\documentclass[twocolumn,showpacs,preprintnumber,amsmath,amssymb,aps]{revtex4}
\usepackage{graphicx}
\begin{document}

\title{  
Moving Vortex 
Phases, Dynamical Symmetry Breaking, and Jamming 
for Vortices in Honeycomb
Pinning Arrays
} 
\author{C. Reichhardt and C.J. Olson Reichhardt} 
\affiliation{ 
Theoretical Division,
Los Alamos National Laboratory, Los Alamos, New Mexico 87545}

\date{\today}
\begin{abstract}
We show using numerical simulations
that vortices in honeycomb pinning arrays can exhibit a remarkable
variety of dynamical phases that are distinct from those found for triangular
and square pinning arrays. In the honeycomb arrays, it is possible
for the interstitial vortices to form dimer or higher
$n$-mer states which have
an additional orientational degree of freedom 
that can lead to the formation of vortex molecular crystals.
For filling fractions where dimer states appear,
a novel dynamical symmetry breaking can occur when the dimers
flow in one of two possible alignment directions.
This leads to transport in the direction transverse to the applied drive. 
We show that
dimerization produces distinct types of moving phases which depend
on the direction of the driving force 
with respect to the pinning lattice symmetry.
When the dimers are driven along certain directions, 
a reorientation of the dimers can produce a jamming phenomenon which
results in a strong enhancement in the critical depinning force.
The jamming can also cause unusual effects such as 
an increase in the
critical depinning force when the size of the pinning sites is reduced. 
\end{abstract}
\maketitle

\vskip2pc

\section{Introduction}

Vortex matter in type-II superconductors has been extensively studied 
as a unique system of many interacting particles
in which nonequilibrium phase transitions can be accessed readily
\cite{Jensen1,Koshelev,Higgins,Giamarchi,Balents,Moon,Pardo,Kes}.
In the absence of driving or quenched disorder, the 
vortex-vortex interactions favor
a triangular crystalline ordering. If the sample contains
sufficiently strong quenched disorder in the form
of randomly placed pinning sites, 
the vortex lattice ordering can be lost as the vortices adjust
their positions to 
accommodate to the pinning landscape 
\cite{Jensen1,Koshelev,Higgins}.
Under an applied drive such as the Lorentz force from a  
current,
the vortices remain
immobile or pinned for low 
driving forces; however, there is a 
threshold applied force above which the vortices begin
to move over the quenched disorder. For strong disorder, the initial 
moving state is
highly inhomogeneous 
with the vortices flowing in meandering and fluctuating channels, and 
there is a coexistence between pinned vortices and 
flowing vortices \cite{Jensen1,Koshelev}.
At higher drives the vortices move more rapidly, the 
effectiveness of the quenched disorder is reduced, and the
fluctuations experienced by the vortices become anisotropic due to 
the directionality of the external drive \cite{Giamarchi}.
The vortex-vortex interactions become more important at the higher drives
when the quenched disorder becomes ineffective,
and a dynamical transition can occur into
a moving smectic state where the vortices regain partial order in one direction
\cite{Balents,Moon,Pardo}. 
Here, the system has
crystalline order in the direction transverse to the vortex motion and
liquid-like order in the direction of vortex motion.
Depending on the dimensionality and the strength of the
pinning, it is also possible for the vortices to reorganize in both directions 
at high drives to form 
a moving anisotropic crystal \cite{Giamarchi,Balents,Moon,Pardo,Kes,Hu}. 
The 
existence of these
different phases and transitions between the phases can be inferred from 
signatures
in transport \cite{Higgins} 
and noise fluctuations \cite{Marley,Maeda}, 
and the moving phases have also been imaged directly using
various techniques \cite{Pardo,Kes}. 

In addition to the naturally occurring randomly placed pinning sites,
it is also possible for artificial pinning sites to be created in a 
periodic structure \cite{Fiory}. 
Recent 
advances in nanostructuring permit the
creation of a wide variety 
of periodic pinning landscapes where 
the periodicity, shape, size and density of the
pinning sites can be well controlled. 
Distinct types of pinning arrays such 
as square \cite{Mosch,Baert,Harada,Rosseel,Rosseel2,Metlushko,Welp,Bending} 
triangular \cite{Schuller,Xiao}, rectangular \cite{Velez,Karapetrov},  
honeycomb \cite{Morgan,Wu}, kagom{\' e} \cite{Morgan}, 
quasicrystalline \cite{Misko1}, and partially ordered \cite{Villegas} 
structures have been created. 
In these arrays
the type of vortex structure that forms is determined by 
whether the vortex lattice 
is commensurate 
with the underlying pinning array.
Commensurate arrangements appear 
at integer multiples of 
the matching field $B_{\phi}$, which is the magnetic 
field at which the vortex density
matches the pinning density, 
and in general, ordered vortex states
occur at matching or rational fractional values of $B/B_{\phi}$ 
\cite{Mosch,Baert,Harada,Periodic1,Libal,Peeters1,Peeters}.   
In samples where only one vortex can be captured by each pinning site,
the vortices that appear above the first matching field 
sit in the interstitial regions between the pinning sites, 
and these interstitial vortices can adopt a variety of crystalline 
configurations
\cite{Baert,Harada,Rosseel,Rosseel2,Welp,Karapetrov,Periodic1,Libal,Peeters}. 

Since a number of distinct ordered and partially ordered vortex states can be 
created in periodic pinning arrays,
a much richer variety of dynamical vortex behaviors occur for periodic
pinning than for random pinning arrays 
\cite{Harada,Rosseel,Rosseel2,Reichhardt,Jensen,Zimanyi,Jensen2,Kolton,Carneiro,Zhu,Look,Benk,Chen,Misko,Kang,Cao,Marconi,Teitel,Rink,Souza}.   
Several of the dynamical phases occur due to the existence of 
highly mobile interstitial vortices which 
channel 
between the pinned vortices  
\cite{Harada,Rosseel2,Zimanyi,Jensen2,Kolton,Zhu,Look,Chen,Misko,Cao,Souza}. 
As a function of applied drive, various types
of moving phases occur, including 
interstitial vortices moving coherently between the pinning sites  
in one-dimensional paths 
\cite{Harada,Rosseel2,Reichhardt,Jensen,Jensen2,Zimanyi,Zhu,Misko} or 
periodically modulated winding paths \cite{Reichhardt,Zimanyi,Kang,Cao}, 
disordered regimes where the vortex motion is liquidlike
\cite{Reichhardt,Zimanyi,Zhu,Misko}, 
and regimes where vortices flow along the pinning rows 
\cite{Reichhardt,Carneiro,Marconi,Teitel,Rink}. 
Other dynamical effects, such
as rectification of mixtures of pinned and interstitial vortices, 
can be realized when the 
periodic pinning arrays are asymmetric \cite{Souza}. 

\begin{figure}
\includegraphics[width=3.5in]{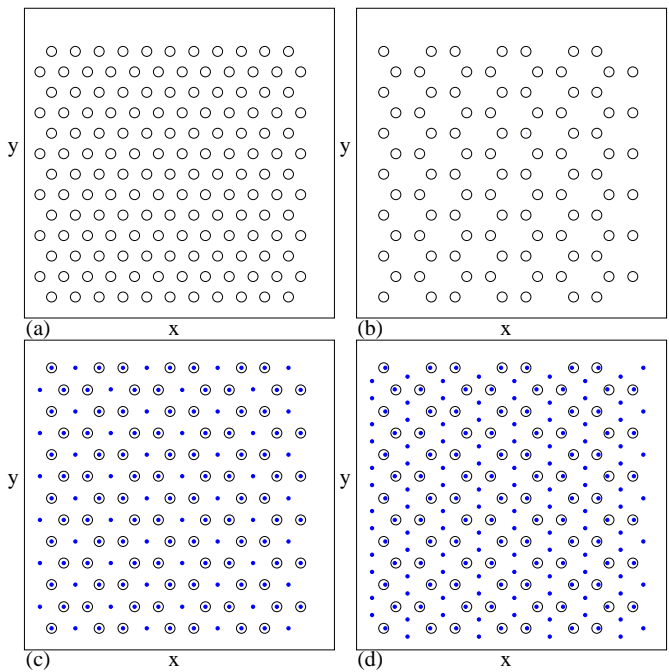}
\caption{
(a) Pinning site locations (open circles) for a triangular pinning array.
(b) Pinning site locations for a honeycomb pinning array 
constructed from the triangular array in (a) by removing $1/3$ of the 
pinning sites.   
(c) The pinning site locations and vortex positions (dots) 
for a honeycomb pinning array at $B/B_{\phi} = 1.5$. 
The overall vortex lattice order is triangular.  
(d) The pinning site locations and vortex positions
for a honeycomb pinning array at $B/B_{\phi} = 2.0$, where 
two vortices are captured at the large interstitial sites and the 
resulting dimers all have the same orientation. 
Here $F_{p} = 0.85$, $R_{p} = 0.35\lambda$, 
and for the honeycomb array $n_{p} = 0.3125/\lambda^2$.
}
\end{figure}

Most of the studies of vortex ordering and dynamics in 
periodic pinning arrays have been performed for square and triangular arrays. 
Experiments with honeycomb and kagom{\' e} pinning arrays revealed
interesting
anomalies in the critical current at nonmatching 
fields which are as pronounced as the anomalies observed
at matching fields in triangular pinning arrays \cite{Morgan,Wu}. 
A honeycomb pinning array is constructed by 
removing every third pinning site from a triangular pinning array, producing
a periodic arrangement of triangular interstitial sites. In 
Figs.~1(a,b) we illustrate a triangular pinning array 
along with the honeycomb pinning array that results after the removal 
of one third of the pinning sites. 
The matching anomalies in the experiments coincide with fields 
$B/B_\phi=m/2$, with $m$ an integer.  At these fields, the 
vortex density would match with the regular triangular pinning array. 
At the matching anomalies for $m>2$, a portion of the vortices
are located in the large interstitial regions 
of the honeycomb lattice \cite{Wu}, as illustrated
in Fig.~1(c) for $B/B_{\phi} = 1.5$. 
The overall vortex lattice structure is
triangular and a strong peak in the depinning force 
occurs at this field \cite{Pinning}.

Recently, we used numerical simulations to demonstrate that vortices 
in honeycomb pinning arrays have a rich equilibrium phase diagram as a 
function of vortex density \cite{Pinning},  with matching anomalies
at integer and half integer matching fields that are  
in agreement with experiments.
The large interstitial sites created by the missing pinning sites can capture
multiple interstitial vortices which form cluster states of $n$ vortices.
For $1.5 \leq B/B_{\phi} < 2.5$, dimer states with $n=2$ form, while
for higher fields
trimer and higher order $n$-mer states form. 
At the integer and half-integer matching fields, the $n$-mer states
can assume a global orientational ordering which 
may be of ferromagnetic or antiferromagnetic type; 
herringbone structures can also form, similar to those observed for 
colloidal particles on periodic substrates 
\cite{Olson,Brunner,Trizac,Frey,Samir} 
and molecules on atomic substrates \cite{Berlinsky}. 
These orientationally ordered states have been 
termed vortex molecular crystals.
Certain vortex molecular crystals 
have ground states that are doubly or triply
degenerate, such as the dimer state 
illustrated in Fig.~1(d) at $B/B_{\phi} = 2.0$ 
where the dimers align in one
of three equivalent directions \cite{Pinning}. 
As the temperature is increased, the $n$-mers undergo a transition from
an ordered state to an orientationally disordered state in which the
$n$-mers rotate randomly but remain confined to the interstitial
pinning sites.
The rotating states have been termed vortex plastic crystals.
At matching 
fields where 
vortex plastic crystals form, the anomalies in the critical current disappear 
\cite{Pinning}. 
The predictions from the simulations are in general agreement with 
the experimental observation of the loss of certain 
higher order matching anomalies at higher temperatures \cite{Wu}.
The formation of $n$-mers that can be aligned along 
degenerate symmetry directions has also been predicted
for kagom{\' e} 
pinning arrays where every other pinning site is removed from every other 
row of a triangular lattice \cite{Pinning,Laguna}.   

The formation of dimer states in the honeycomb pinning 
array leads to a variety of novel dynamical phases, including
a spontaneous dynamical symmetry breaking effect 
in which the moving vortices organize into one of two equivalent states 
which have a component of translation perpendicular to the applied drive in
either the positive or negative direction
\cite{Transverse}. 
The transverse response appears when the external driving force is 
applied halfway between  
the two symmetric directions of aligned dimer motion.
The dynamical symmetry breaking
occurs when the equilibrium ground
states have no global symmetry breaking.
At $B/B_{\phi} = 2.0$, 
the ground state is symmetry broken and the dynamical moving state 
has the same broken symmetry as the ground state. 
For incommensurate fillings, when the dimer alignment is disrupted, there
is no global symmetry breaking in the ground state, and instead a
dynamical symmetry breaking occurs due to the applied drive. 

In this work we map the dynamical phase diagram 
for vortices in honeycomb arrays. We focus on the states 
$1.5 < B/B_{\phi} < 2.5$     
to understand where 
dynamical symmetry breaking occurs and to examine what other 
types of moving phases are possible. 
We study how the 
dynamical phases change for driving along different axes of the
pinning lattice. 
We find that very different kinds of dynamics occur when the driving
direction is varied, 
and that the value of the depinning threshold is strongly directionally
dependent.
We also find that a novel jamming phenomenon 
can occur due to the formation of the
dimer states. For certain directions of drive, 
the dimers are anti-aligned with the drive, 
causing the dimers to become blocked in the interstitial regions.  

Although our results are specifically for vortices in type-II superconductors,
the general features of this work should also be relevant for other
interacting particle systems where a periodic substrate is present.  
Examples of such systems include vortices on periodic substrates 
in Bose-Einstein condensates (BEC), 
where different kinds of
crystalline phases can occur which depend on the strength of the substrate 
\cite{Bigelow,Tung}.
It should be possible to observe different types of vortex
flow states in BEC systems \cite{Dynamics}.
Our results are also relevant for
colloids on periodic substrates,     
where an orientational ordering of colloidal molecular crystals 
occurs which 
is very similar to that of the vortex molecular 
crystal states \cite{Olson,Brunner,Trizac,Frey,Samir,Bechinger}.  
Other related systems include charged balls on periodic substrates 
\cite{Coupier}
and models of sliding friction \cite{Braun}.  

\section{Simulation} 

We use the same simulation employed in the previous study of vortex
equilibrium states in honeycomb pinning arrays \cite{Pinning}. 
We consider a 2D system of dimensions $L_x=L$ and $L_y=L$  
with periodic boundary conditions in the $x$ and $y$ directions.
The sample contains $N_{v}$ vortices, 
giving a vortex density of $n_{v} = N_{v}/L^2$ which is proportional to the
external magnetic field.
In addition, there are $N_{p}$ pinning sites placed in a honeycomb arrangement
with a pinning density of $n_{p}=N_p/L^2$.
The field at which the number of vortices equals the number 
of pinning sites is defined to be the matching field $B_{\phi}$. 

The dynamics of vortex $i$ located at position ${\bf R}_i$ is governed 
by the following
overdamped equation of motion:
\begin{equation}
\eta\frac{d {\bf R}_{i}}{dt} =  {\bf F}^{vv}_{i} + {\bf F}^{vp}_{i} 
+  {\bf F}_{D}  + {\bf F}_i^T.
\end{equation} 
Here the damping constant is $\eta = \phi^{2}_{0}d/2\pi \xi^2 \rho_{N}$,
where $d$ is the thickness of the superconducting sample, 
$\xi$ is the superconducting coherence length,
$\rho_{N}$ is the normal state resistivity of the material, 
and $\phi_{0} = h/2e$ is the
elementary flux quantum. 
The vortex-vortex interaction force is
\begin{equation} 
{\bf F}^{vv}_{i} = \sum^{N_{v}}_{j \neq i}f_{0}K_{1}\left(\frac{R_{ij}}{\lambda}\right){\bf {\hat R}}_{ij}
\end{equation} 
where $K_{1}$ is the modified Bessel function, 
$\lambda$ is the London penetration
depth, $f_{0} = \phi^{2}_{0}/(2\pi\mu_0\lambda^{3})$, 
$R_{ij}=|{\bf R}_i-{\bf R}_j|$ is the distance between
vortex $i$ and vortex $j$, and the unit vector 
${\bf {\hat R}}_{ij} = ({\bf R}_{i} - {\bf R}_{j})/R_{ij}$.
In this work all length scales are measured in units of $\lambda$ and 
forces in units of $f_{o}$.
The vortex vortex interaction decreases sufficiently rapidly at large distances
that a long range cutoff is placed on the interaction force
at $R_{ij} = 6\lambda$ 
to permit more efficient computation times. 
We have found that the cutoff does not affect 
the results for the fields and forces we consider here.

The pinning force ${\bf F}^{vp}_i$ originates from
individual nonoverlapping attractive parabolic 
traps of radius $R_{p}$ which have a maximum strength of $F_{p}$. 
In this work we consider the limit where only   
one vortex can be captured per pinning site, with the majority of the
results obtained for $R_{p} = 0.35\lambda$. 
The exact form of the pinning force is:
\begin{equation}
{\bf F}^{vp}_{i} = -\sum_{k=1}^{N_p}f_0\left(\frac{F_{p}}{R_{p}}\right)
R_{ik}^{(p)}\Theta\left(\frac{R_{p} - R_{ik}^{(p)}}{\lambda}\right){\bf {\hat R}}^{(p)}_{ik} .
\end{equation}
Here, $R_{ik}^{(p)}=|{\bf R}_{i} - {\bf R}_{k}^{(p)}|$,
$R^{(p)}_{k}$ is the location of pinning site $k$,
the unit vector 
${\bf {\hat R}}^{(p)}_{ik} = ({\bf R}_{i} - {\bf R}_{k}^{(p)})/R_{ik}^{(p)}$,
and $\Theta$ is the Heaviside step function. 

The external drive ${\bf F}_{D}=F_Df_0{\bf {\hat F}_D}$ 
represents the Lorentz force from an applied current 
${\bf J}\times{\bf B}$ which is perpendicular to the
driving force and is applied uniformly to all the vortices. 
We apply the drive at various angles to the symmetry axes of the honeycomb
pinning array.
The thermal force ${\bf F}_i^T$ originates from Langevin kicks with the
properties $\langle {\bf F}_i^T\rangle=0$ and
$\langle {\bf F}_i^T(t){\bf F}_j^T(t^\prime)\rangle=
2\eta k_BT \delta_{ij}\delta(t-t^\prime)$.
Unless otherwise noted, the thermal force is set to zero.
The initial vortex configurations are obtained by 
simulated annealing,
and the external force is then applied gradually in increments
of $\Delta F_{D}=0.0002$ every 1000 simulation time steps. 
For the range of pinning forces used in this work, 
we find that this force ramp
rate is sufficiently slow that transients in the vortex dynamics do
not affect the overall velocity-force curves. 
We obtain the velocity-force curves by summing the 
velocities in  the $x$ (longitudinal) direction, 
$\langle V_{x}\rangle = N_{v}^{-1}\sum^{{N}_{v}}_{i=1} {\bf v}_{i}\cdot 
{\bf {\hat x}}$, and the 
$y$ (transverse) direction, 
$\langle V_{y}\rangle = N_{v}^{-1}\sum^{N_{v}}_{i}{\bf v}_{i}\cdot 
{\bf {\hat y}}$, where
${\bf v}_{i} = d{\bf R}_{i}/dt$.   
In Fig.~1(c,d) we illustrate the pinning sites and vortex configurations
after simulated annealing for $B/B_{\phi} = 1.5$ [Fig.~1(c)] 
and $2.0$ [Fig.~1(d)].  
Here $L_x=L_y = 24\lambda$ and $n_{p} = 0.3125/\lambda^2$.
In our previous work, Ref.~\cite{Transverse}, the drive was applied along
the $x$-direction for the geometry in Fig.~1. 

\section{Dynamics and Transverse Response For Driving in the Longitudinal Direction} 

\begin{figure}
\includegraphics[width=3.5in]{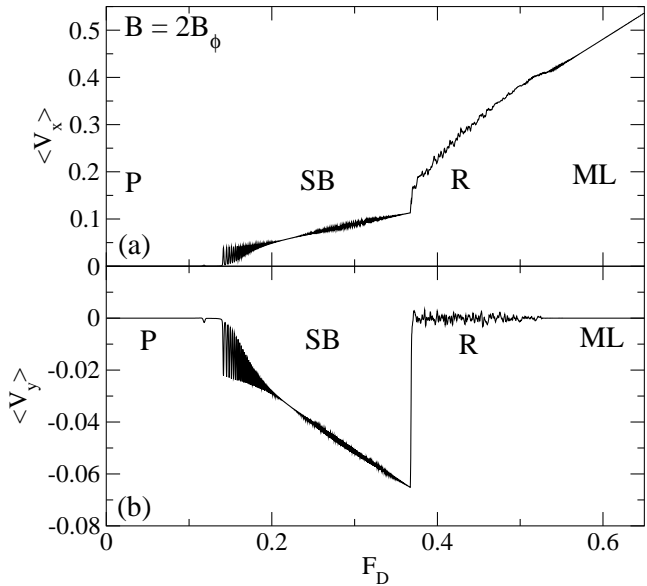}
\caption{
(a) The average velocity in the $x$-direction 
$\langle V_{x}\rangle$ vs external driving force $F_{D}$ for the
honeycomb pinning array from Fig.~1(d) at $B/B_{\phi} = 2.0$ 
with ${\bf F}_{D}=F_D{\bf {\hat x}}$.
(b) The corresponding average velocity in the $y$-direction 
$\langle V_{y}\rangle$ vs $F_{D}$.    
We observe four phases: the initial pinned phase (P), 
a symmetry broken phase (SB), a random phase (R),
and a moving locked phase (ML). 
}
\end{figure}

\begin{figure}
\includegraphics[width=3.5in]{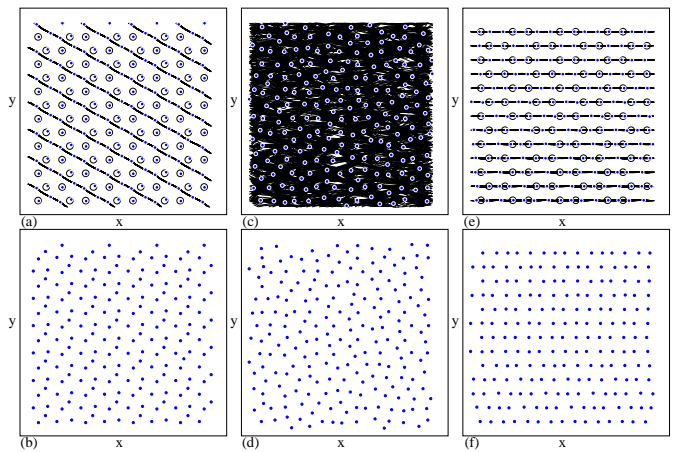}
\caption{ 
The dynamics of the three moving phases from Fig.~2 for the honeycomb
pinning array at $B/B_\phi=2.0$ with ${\bf F}_D=F_D{\bf {\hat x}}$.
The vortex positions (filled circles), 
pinning site locations (open circles), and
vortex trajectories (black lines) are shown in an $18\lambda \times 18\lambda$
portion of the sample.
(a) In the symmetry broken (SB) phase at $F_{D} = 0.25$, 
the interstitial vortices move along a $-30^{\circ}$ angle to the $x$-axis
while the vortices at the pinning sites remain immobile. 
(b) Vortex positions only in the SB phase at $F_D=0.25$, 
showing the ordering present in the vortex lattice structure.
(c) In the random (R) phase at $F_{D} = 0.42$,  the vortex motion is
highly disordered with vortices pinning and repinning at random. 
(d) Vortex positions only in the R phase at $F_D=0.42$ indicate that
the vortex lattice is disordered.
(e) In the moving locked (ML) phase
at $F_{D} = 0.65$, all the
vortices channel along the pinning sites. 
(f) Vortex positions only in the ML phase at $F_D=0.65$
reveal an anisotropic vortex lattice structure 
with different numbers of vortices in each row.   
}
\end{figure}

We first consider the case for driving in the $x$ or longitudinal direction,
${\bf F_D}=F_D{\bf {\hat x}}$,
for the system shown in Fig.~1(d) with 
$B/B_{\phi} = 2.0$, $R_{p} =0.35\lambda$, and  $F_{p} = 0.85$. 
In Figs.~2(a,b) we plot $\langle V_{x}\rangle$ and 
$\langle V_{y}\rangle$ versus $F_{D}$.
At this filling there are four distinct dynamical phases,
with the pinned (P) phase occurring at low $F_{D}$. 
The depinning threshold $F_{c}$ occurs
near $F_{D} = 0.14$ when the interstitial vortices become depinned.
For a system with random pinning and ${\bf F}_{D}=F_{D}{\bf {\hat x}}$,
there would be no transverse velocity response; 
the system would have $\langle V_{y}\rangle = 0$ 
and only $\langle V_{x}\rangle$ would be finite. 
In contrast, for the honeycomb pinning array there is 
a finite velocity both
in the positive $x$ direction and in either 
the $+y$ or $-y$ direction. 
In Fig.~2(b) the transverse response $\langle V_y\rangle$
is negative, indicating that the
vortices are moving at a negative angle to the $x$ axis 
for $0.14 < F_{D} < 0.37$.  
Figure 3(a) illustrates the vortex motion at $F_D=0.25$, where
the vortices flow in one-dimensional paths oriented at $-30^{\circ}$ 
to the $x$ axis. 
In Fig.~3(b) a snapshot of the vortex positions
shows that the vortex lattice remains ordered in the moving phase, 
indicating that the vortices
are flowing in a coherent manner.   
We term the phase shown in Fig.~3(a) the symmetry broken 
(SB) phase,
since the flow can be tilted in either the positive or negative $y$-direction. 

At $B/B_{\phi} = 2.0$ and $F_{D} = 0$, 
the interstitial vortices form an aligned dimer 
configuration with a three-fold degenerate ground state 
in which the dimers can be oriented along the $y$-direction, 
as in Fig.~1(d), or along $+30^{\circ}$ or $-30^{\circ}$ to the $x$-direction, as
shown in previous work \cite{Transverse}.
When a driving force is applied to the $+30^{\circ}$ or $-30^{\circ}$ ground
states, 
the vortices depin and flow along $+30^{\circ}$ or $-30^{\circ}$, respectively.
In these cases, the symmetry breaking in the moving state
is not dynamical in nature 
but reflects the symmetry breaking within the ground state.  
If the dimers are initially aligned along the $y$-direction
in the ground state, an applied drive 
induces an instability in the pinned phase and
causes the dimers to rotate into the $+30^{\circ}$ or $-30^{\circ}$ directions,
as we discuss in further detail below.  In this case the symmetry breaking
is dynamical in origin.

\begin{figure}
\includegraphics[width=3.5in]{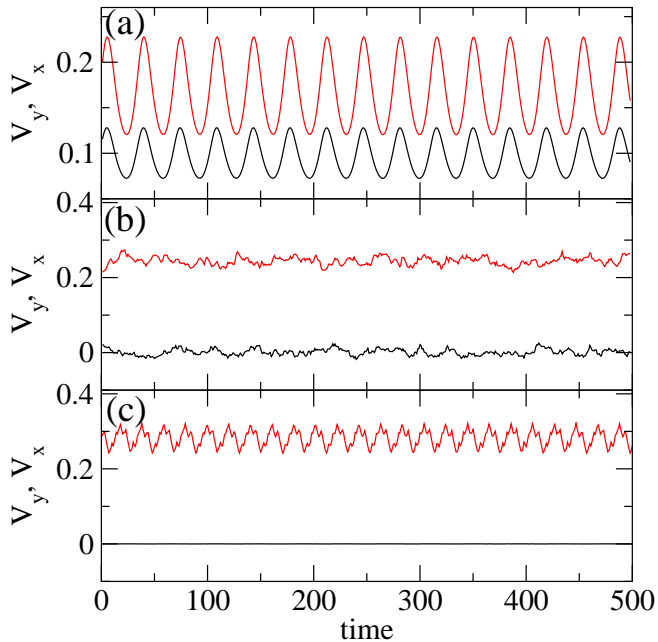}
\caption{
Time traces of vortex velocity at fixed $F_D$.
Upper curves: $V_x(t)$; lower curves: $V_y(t)$.
(a) 
The symmetry broken (SB) phase at $F_{D} = 0.25$ from Fig.~3(a,b). 
Here pronounced oscillations occur in both $V_x$ and $V_y$
as the vortices move in a coherent fashion. 
(b) The random (R) phase at $F_{D} = 0.42$ from Fig.~3(c,d). In this case
the transverse motion is lost and $\langle V_{y}\rangle = 0$. 
Additionally, there are no correlated oscillations. 
(c) The moving locked (ML) phase at $F_{D} = 0.65$ from Fig.~3(e,f). 
$V_x$ has been shifted down for clarity. 
There is a weak oscillation in $V_x$ due to the periodic substrate. 
Since the flow is strictly one-dimensional, as shown in Fig.~3(e), 
there are no fluctuations
in $V_y$.
}
\end{figure}

In Fig.~2(a,b)
we find pronounced oscillations in both 
$\langle V_{x}\rangle$ and $\langle V_{y}\rangle$
just above the depinning threshold $F_c=0.14$. 
These oscillations are not intrinsic features but are due to the fact 
that at $B/B_{\phi} = 2.0$ the interstitial vortex lattice is perfectly 
ordered, so the interstitial vortices move in 
a coherent fashion as shown in Fig.~3(a). 
At depinning, the interstitial vortices are 
slowly moving through a periodic potential created by vortices 
that remain trapped at the pinning sites. This periodic 
potential causes the moving interstitial vortices to develop
an oscillating velocity.
In Fig.~4(a), the instantaneous time 
traces of the vortex velocity $V_x$ and $V_y$ at constant $F_{D} = 0.25$ 
show strong velocity oscillations. 
The oscillations are also visible in Fig.~2 at low drives due to our choice
of averaging time spent at each value of the driving current.  If the averaging
time is increased, the oscillations in Fig.~2 disappear.  We note that the
locations of the boundaries between the different phases are not affected by
the value of the velocity averaging time.
At incommensurate fields, 
there is enough dispersion in the velocity of the moving interstitial
vortices that the coherent velocity oscillations are 
no longer distinguishable.

As $F_{D}$ increases, the net vortex velocity 
in the SB phase  
increases linearly until $F_{D} = 0.365$, where there is an abrupt increase 
in $\langle V_{x}\rangle$.  Fig.~2(a,b) shows that this increase 
coincides with a jump in $\langle V_{y}\rangle$ to a zero average, 
indicating that the vortices are moving only in
the $x$-direction on average.
In Fig.~3(c) we illustrate the disordered vortex trajectories that occur
in this phase at $F_{D} = 0.42$.
The vortices are continually depinning and being repinned, 
and the order in the vortex lattice is lost, as shown in Fig.~3(d). 
We term this the random (R) phase.  It resembles random dynamical phases that
have previously been observed for 
vortices in square pinning arrays 
when the interstitial vortices begin to depin
vortices from the pinning sites \cite{Periodic1}.
Figure 2 shows that
there are pronounced random fluctuations in  
$\langle V_{x}\rangle$ and $\langle V_{y}\rangle$ in phase R, and also
that $\langle V_{x}\rangle $ does not increase linearly 
with $F_D$ but has a curvature
consistent with $ V_{x} = (F_{D} - F^{R}_{c})^{1/2}$, 
where $F^{R}_{c}=0.365$ is the threshold value for
the SB-R transition. 
In the SB phase, the number of moving vortices is constant and is equal to
the number of interstitial vortices, while in the R
phase the number of moving vortices increases with $F_D$.     

At $F_{D} = 0.53$, the system 
organizes into a one-dimensional flowing state where the vortex motion 
is locked along the pinning rows, as shown in Fig.~3(e,f) for $F_{D} = 0.65$. 
The onset of this phase also coincides with the decrease of fluctuations 
in $\langle V_{x}\rangle$ and the loss of fluctuations in 
$V_y$, as shown in Fig.~4(c).
For $F_{D} > 0.53$, 
all of the vortices are mobile and 
Fig.~2(a) illustrates that the $\langle V_{x}\rangle$ versus $F_{D}$ 
curve becomes linear again.              
We term this the moving locked (ML) phase, since 
the vortex motion is effectively locked along the pinning sites. 
When the vortices are rapidly moving, 
the pinning sites have the same effect as a flashing one-dimensional 
trough that 
channels the vortices \cite{Reichhardt,Carneiro}.     
The vortices assume a smectic structure in the ML phase, 
since different rows have different numbers
of vortices, 
resulting in the formation of
aligned dislocations. The ML phase
is essentially the same state found in square pinning arrays at high drives
when $B/B_{\phi} > 1.0$ \cite{Reichhardt}. 

In previous studies of square pinning arrays with strong pinning, 
the initial motion of the vortices for $B/B_{\phi} > 1.0$ 
occurred in the form of one-dimensional channels 
between the vortices trapped at the pinning sites \cite{Reichhardt}. 
In the honeycomb pinning array, similar flow occurs in the SB phase
as shown in Fig.~3(a).
For $B/B_{\phi} < 1.5$ in the honeycomb array, 
the initial interstitial flow for depinning in the 
$x$ direction occurs via the flow of individual vortices 
in a zig-zag pattern around the pinned vortices.
Since there is no dimer ordering for these fillings, 
no transverse response occurs for $B/B_{\phi} < 1.5$. 
For $B/B_{\phi} \ge 1.5$, the interstitial vortices
begin to form dimer states when two interstitial vortices are captured
in a single large interstitial site.
The dimers can lower their 
orientational energy by aligning with each other in both
the ground state and the moving states. 
Dimers can only remain aligned in the moving state if they are channeling
along one of the symmetry axes of the pinning lattice. 
If the dimers were to move strictly in  
the $x$-direction, they
would be forced directly into the pinned vortex in the 
pinning site to the right of each large interstitial site. 
This would destabilize the rodlike dimers.  Instead, the dimers maintain
their integrity by moving along $\pm 30^{\circ}$ to the $x$-axis.
Within the moving state, if one of the dimers were to move along $+30^\circ$
while the remaining dimers were moving along $-30^\circ$,
the two interstitial vortices comprising the first dimer would 
be forced close together, destabilizing the dimer state due to the repulsive
vortex-vortex interactions.  Instead, all of the dimers move in the
same direction.

The SB-R transition occurs when
the combined forces on the pinned vortices from the external drive and the 
moving dimers are strong enough to depin the pinned vortices.
At the closest approach in the $x$ direction between a dimer
and a pinned vortex, 
the frontmost dimer vortex 
is a distance $a_{0}/2$ from the pinned vortex and 
the rear dimer vortex is a distance $3a_{0}$ from the pinned vortex,
where $a_{o}$ is the 
lattice constant of the undiluted triangular pinning lattice.
In addition to the force from the dimerized vortices,
the pinned vortex experiences an opposing force
from the neighboring pinned vortex a distance $a_0$ away. 
In a simple approximation, 
the driving force needed to depin a vortex at a pinning site is  thus
$F_{D} = F_{p} - [(K_{1}(a_{0}/2) + K_{1}(3a_{0}/2)) - K_{1}(a_{0})]$. 
Setting $F_{p} = 0.85$ gives $F_{D} = 0.41$, close to 
the value of $F_{D} = 0.37$ for the SB-R transition in Fig.~2. 
Once the pinned vortices depin, the system enters the random (R) phase,
and since $F_{D}$ is still considerably less than $F_{p}$, it is 
possible for vortices to be pinned temporarily in phase R. 

Studies  
of square pinning arrays have shown
that after the onset of a random dynamical phase, 
the vortices can organize 
into a more ordered phase of solitonlike pulse
motion along the pinning rows, followed by a phase in which all of the
vortices channel along the pinning rows \cite{Reichhardt}. 
At the transition to the one-dimensional pulse like motion, 
a larger fraction of the vortices are
pinned compared to the random phase,  
so a drop in $\langle V_{x}\rangle$ with increasing $F_{D}$ 
occurs, giving a negative differential conductivity.
In the honeycomb pinning arrays for the parameters we have chosen 
here, we do not observe  
one-dimensional pulse motion or negative differential conductivity for 
driving along the $x$-direction. 
For the one-dimensional pulse motion or the ML phase motion seen 
in Fig.~3(e,f) to occur, the vortices must be moving at 
a sufficiently high velocity for the pinning sites 
to act like a flashing trough. 
When the vortices move along the pinning rows, the vortex lattice structure
adopts a highly anisotropic configuration which would be unstable at
$F_D=0$.
During the period of time when a vortex passes through a pinning site, the
vortex is pulled toward the center of the pinning row, which stabilizes the
one-dimensional motion.
When the vortex is moving between the pinning sites, it can drift 
away from the one-dimensional path until 
it encounters another pinning site. 
In Ref.~\cite{Reichhardt}, it was shown that for
square pinning arrays, increasing the pinning radius
$R_p$ stabilized the one-dimensional flow down to lower values of $F_D$. 
In the honeycomb pinning array, the one-dimensional flow is less stable due
to the fact that the vortices must move over the much wider large interstitial
site, giving the vortices more time to drift away from the pinning row.
Since this means that a larger value of $F_D$ is required to stabilize the
one-dimensional motion, it 
should be more difficult in general to observe the
onset of one-dimensional solitonlike motion 
or negative differential conductivity in the honeycomb pinning arrays
than in the square pinning arrays.   

\begin{figure}
\includegraphics[width=3.5in]{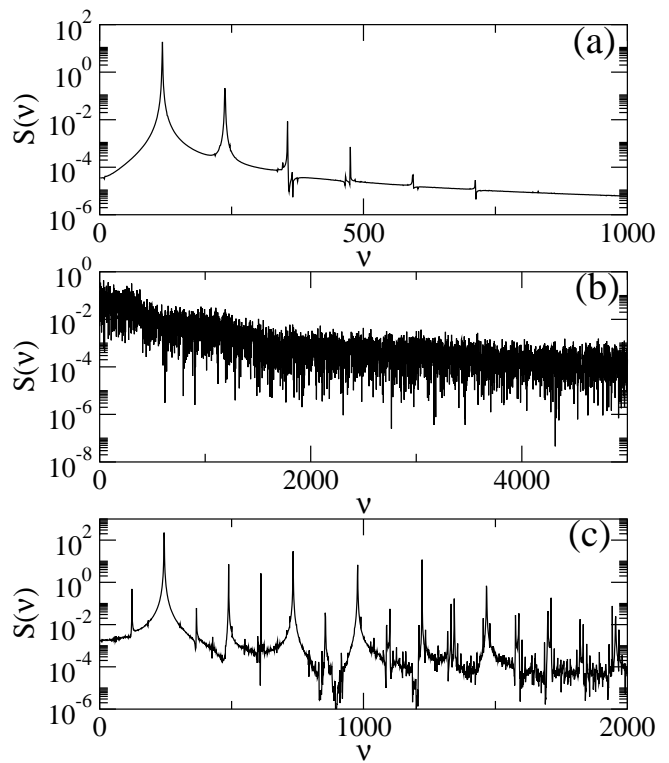}
\caption{
The power spectra $S(\nu)$ of the $x$ component of the velocity 
$V_x(t)$ for the three phases in Fig.~4.
(a) The SB phase at $F_D=0.25$ shows a pronounced narrow band 
noise signature. (b) The R phase at $F_D=0.42$ has a
broad band noise signature. 
(c) In the ML phase at $F_D=0.65$,
a number of different frequencies are present 
due to the fact that different rows 
of the vortices move at different velocities.  
}
\end{figure}

\subsection{Fluctuations and Noise Characteristics} 

In order to characterize the 
moving phases more quantitatively, 
in Fig.~4 we show time traces $V_x(t)$ and $V_y(t)$ 
of the vortex velocities at fixed 
$F_{D}$ for the different phases for the system in 
Fig.~2. 
In the symmetry broken (SB) phase at $F_D=0.25$, shown in Fig.~4(a),
$V_x$ is greater than $|V_y|$ by
$\tan(30^{\circ})$ or about $1.7$.
Here both components of the velocity 
show a pronounced oscillation which arises when the
interstitial vortices move in a coherent fashion over  
the periodic potential substrate 
created by the immobile vortices in the pinning sites.
In Fig.~5(a), we plot the corresponding power spectrum 
$S(\nu)$
of $V_{x}$ obtained from 
\begin{equation} 
S(\nu) = \left |\int V_x(t)e^{-2\pi i\nu t}dt\right|^2 .
\end{equation} 
There is a pronounced peak in $S(\nu)$ at the frequency of the
velocity oscillation in the SB phase, indicating that
mode locking effects could appear at 
$B/B_{\phi} = 2.0$ when the symmetry breaking flow occurs.  
In square pinning arrays, experiments \cite{Rosseel2} 
and simulations \cite{Jensen2} revealed
Shapiro step-like mode locking of 
interstitial moving vortices at $B/B_{\phi} = 2.0$.
In the honeycomb lattice, since there is also a strong 
oscillation in $V_y$ in the SB phase,
we expect that transverse mode locking could occur if an
additional ac drive is applied in the 
$y$-direction.  Such mode locking would appear as steps in both 
$\langle V_{x}\rangle$  
and $\langle V_{y}\rangle$ versus $F_{D}$ in the SB phase. 
Transverse phase locking, which produces steps that are distinct from
Shapiro steps,  has been observed for the 
motion of vortices in square arrays
\cite{Kolton}.  
In general, if the vortices already have an 
intrinsic velocity oscillation in the transverse
direction, then pronounced transverse phase locking is possible.
           
In Fig.~4(b) we plot the time trace of $V_{x}$ and $V_{y}$ 
for the random (R) phase at $F_{D} = 0.42$. 
In this case $\langle V_{y}\rangle = 0$, 
and although both $V_x$ and $V_y$ show fluctuations,
no oscillations or washboard frequencies appear. 
In Fig.~5(b) we show the corresponding $S(\nu)$ for $V_{x}$, where 
we find a broad band noise feature 
consistent with disordered plastic flow \cite{Higgins,Moon,Maeda}.
Since there are no coherent velocity oscillations, 
mode locking should be absent in the random (R) phase.

In Fig.~4(c) we plot $V_x$ and $V_y$ in
the ML phase at $F_{D} = 0.65$, where $V_{x}$ has been shifted
down by a factor of 3 for clarity. 
There are no visible fluctuations in $V_{y}$ due to the one-dimensional
nature of the flow, but 
there are small periodic oscillations 
in $V_{x}$ generated by the motion of the vortices 
over the periodic substrate. 
Due to the fact that different one-dimensional rows contain different numbers
of vortices, producing dispersion in the vortex velocities, 
the oscillation in $V_x$ is not as pronounced as in the SB phase.
The corresponding 
power spectrum in Fig.~5(c) contains a rich variety of peaks 
due to the wide range of frequencies present in this phase.
The main peak is smaller in magnitude than that found for the SB phase. 
As $F_{D}$ increases, the frequency at which the first peak occurs
also increases.  
It should be possible to generate phase locking in the ML phase; however, it
would likely not be as pronounced 
as in the SB phase. 
These results suggest that noise fluctuations 
can be a useful technique for exploring the 
presence of different dynamical phases in periodic pinning arrays.  

\begin{figure}
\includegraphics[width=3.5in]{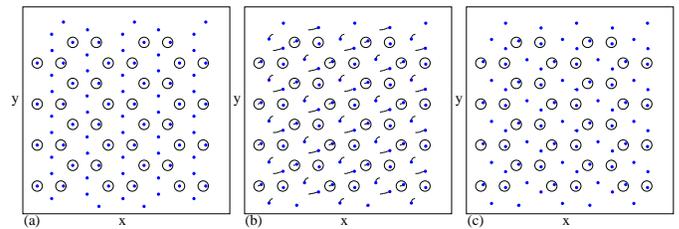}
\caption{
Vortex positions (filled circles), pinning site locations (open circles), and
trajectories (black lines) for a system with $B/B_{\phi} = 2.0$ 
and $F_{p} = 0.85$ which started in the $y$-aligned ground state.  
(a) The pinned phase at $F_{D} = 0.09$. Here the dimers and the vortices in
the pinning sites have all shifted slightly to the right compared to the
ground state due to the applied drive.
(b) A rotational instability occurs at $F_D \approx 0.11$, when
the vortices move in a manner that allows the 
dimers to align along $-30^\circ$ to the $x$-axis.
There is also a small shift of the vortices in the pinning sites. 
(c) The pinned state at $F_{D} = 0.12$ 
where the dimers are aligned in the new $-30^\circ$ direction.    
}
\end{figure}

\begin{figure}
\includegraphics[width=3.5in]{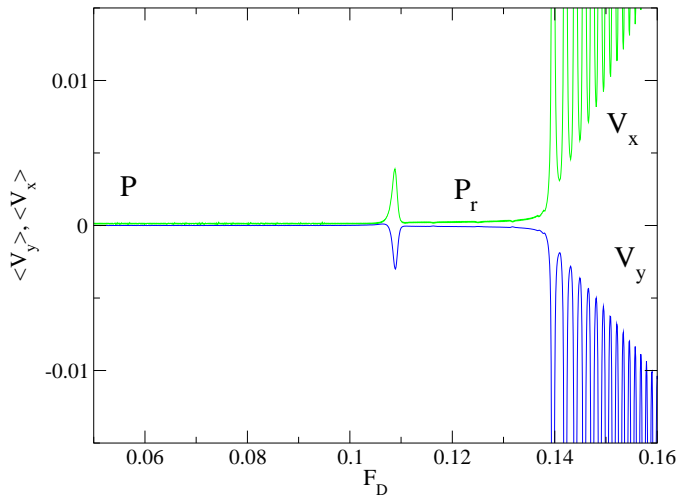}
\caption{
The velocity $\langle V_x\rangle$ (upper curve) and 
$\langle V_y\rangle$ (lower curve) vs $F_D$ for the system in 
Fig.~6. The rotational instability seen in Fig.~6(b) 
appears as a positive peak in $\langle V_{x}\rangle$
and a negative peak in $\langle V_{y}\rangle$ just above $F_{D} = 0.11$. 
The system remains pinned until around $F_{D} = 0.14$.     
}
\end{figure}

\subsection{Dynamical Symmetry Breaking in the Pinned Phase}  

As previously noted, the ground state at $B/B_{\phi} = 2.0$ 
is three-fold degenerate. When the
dimers are aligned at either $+30^\circ$ or $-30^{\circ}$ to the $x$-axis in
the ground state,
the subsequent SB flow is aligned in the same direction as the ground state.
It is also possible for the dimers to align in the $y$-direction, as shown 
in Fig.~6(a).  At $F_{D} = 0.09$,
the dimers and the vortices in the 
pinning sites are shifted slightly to the right due to the applied drive.
As $F_{D}$ is further increased,
a symmetry breaking transition occurs within the pinned phase. 
For $F_{D} < 0.11$ the dimers
remain aligned the $y$-direction; however, at $F_{D} \approx 0.11$, 
the rotational instability illustrated 
in Fig.~6(b) occurs.
The dimers rotate 
in such a way that they end up aligned in the $-30^{\circ}$ direction. 
The interstitial vortex at the bottom of the dimer moves 
in the $+x$ direction and by a smaller amount in the
$+y$ direction, while the vortex at the top of the dimer moves in the 
$-y$ direction and by a
smaller amount in the $-x$ direction. 
There is also a slight shift of the vortices
in the pinning sites that are closest to the bottom of each dimer. 
In Fig.~6(c) the rotation process is completed and the dimers are 
aligned in a new direction, $-30^\circ$.
The vortices remain pinned until $F_{D} =  0.14$, 
at which point the system enters the SB phase.  
At finite temperatures, the 
dimer realignment occurs at even lower values of $F_{D}$. 
The rearrangement can also be 
observed as a jump in $\langle V_{x}\rangle$ and $\langle V_{y}\rangle$ 
as shown in Fig.~7, where there is a positive spike in $\langle V_x\rangle$
and a negative spike in $\langle V_y \rangle$
near $F_{D} = 0.11$, in agreement with the motion 
shown in Fig.~6(b). 
We term this  a dimer  polarization effect 
since the driving force induces an alignment of
the dimers. 
In runs with slightly different initial conditions, the dimers may align
along the $+30^\circ$ direction with 50\% probability.

\begin{figure}
\includegraphics[width=3.5in]{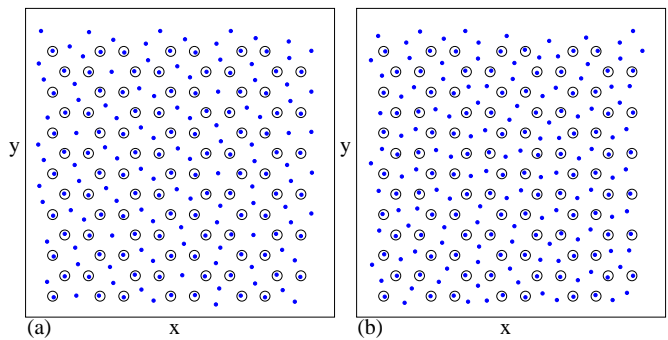}
\caption{
Vortex positions (filled circles) and pinning site locations 
(open circles) for the system in Fig.~2 at
(a) $B/B_{\phi} = 1.91$ and (b) $ B/B_{\phi} = 2.08$. 
At these fillings, the long range orientational ordering 
of the dimer state is lost.  
}
\end{figure}

\subsection{Dynamics for $1.5 \leq B/B_{\phi} <  2.5$}  

We next consider the effect of changing the vortex density for fillings 
where interstitial dimers are present 
and the SB phase occurs. 
In Fig.~8(a) we illustrate the vortex positions 
for $B/B_{\phi} = 1.91$, where a mixture of monomers and dimers appear in 
the large interstitial sites.
At this filling, the overall orientational ordering of the dimers is lost in the
ground state, and the dimers are oriented only in local patches.
For $B/B_{\phi} > 2.0$, a mixture of interstitial dimers and trimers 
is present, as shown in Fig.~8(b) 
for $B/B_{\phi} = 2.08$, and the orientational ordering is again lost. 
In Ref.~\cite{Transverse}
it was shown that the SB state still occurs at incommensurate fields 
as long as some dimers are present.
If $F_D$ is suddenly increased from zero to a finite value at which only the
interstitial vortices depin, the moving state for the incommensurate fields
organizes into a dynamically symmetry broken state where all of the 
dimers flow along $+30^\circ$ or $-30^\circ$.
At the incommensurate fields, only the dimers undergo dynamical
symmetry breaking; the monomers and trimers
continue to move in the direction of the drive,
with some fluctuations in the transverse direction.        

\begin{figure}
\includegraphics[width=3.5in]{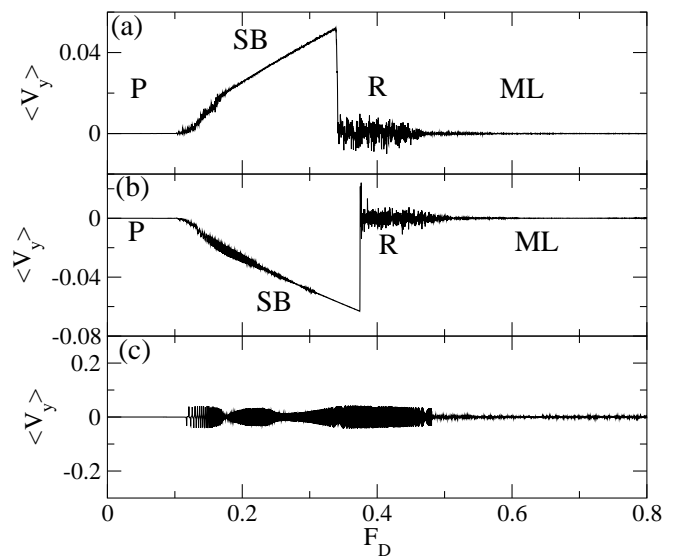}
\caption{
$\langle V_y\rangle$ versus $F_{D}$ for (a) $B/B_{\phi} = 1.89$, 
(b) $B/B_{\phi} = 1.94$, and (c) $B/B_{\phi} = 2.5$ 
for a system with the same parameters as in Fig.~2.
P: pinned phase; SB: symmetry broken phase; R: random phase;
ML: moving locked phase.  
}
\end{figure}

In Fig.~9 we plot $\langle V_{y}\rangle$ for the system in Fig.~2 
at $B/B_{\phi}  = 1.89$, $1.94$, and $2.5$. 
In Figs.~9(a,b), the same four phases described above
are labeled.
The SB phase has opposite sign in Fig. 9(a) and Fig. 9(b);
the dynamical symmetry breaking can occur in either direction since there
is no symmetry breaking in the ground state. 
If slightly different initial conditions are used, 
such as by changing the initial annealing procedure, the
dynamical symmetry breaking has equal probability to 
occur in the positive or negative direction, 
as shown previously \cite{Transverse}. 
In Figs.~9(a) and (b) the initial portion of the 
SB phase has fluctuations in $\langle V_{y}\rangle$ 
due to the fact that we are increasing $F_{D}$ at a finite
rate and there is a transient time for the moving state to 
fully organize into the SB state, as studied previously 
\cite{Transverse}.
The transient time increases as $|B/B_\phi-2.0|$ increases. 
If we decrease $\Delta F_{D}$, the fluctuations at depinning are 
reduced; however, the boundary between the phases
does not shift. 

In Fig.~9(c) at $B/B_\phi=2.5$, $\langle V_{y}\rangle  = 0$ 
since there are only trimer states present. The large oscillations
in $\langle V_{y}\rangle$ occur when the system forms a completely ordered
trimer ground state \cite{Pinning} and the vortex motion is highly coherent,
similar to the effect seen in Fig.~2.
For the rate at which we sample 
and average $\langle V_{y}\rangle$ versus $F_{D}$, 
the periodic fluctuating vortex velocity is visible.
For $F_{D} > 0.5$ the 
system enters a partially moving locked phase where a 
portion of the vortices move along the pinning rows.
There are, however, too many vortices to form straight one-dimensional 
chains of the type shown in Fig.~3(e) for $B/B_{\phi} = 2.0$. 
A buckling instability of the chains occurs since the amount of anisotropy
that would occur if one-dimensional chains formed is too large for the
vortex lattice to sustain.
Instead, a partially moving locked (PML) phase forms with a disordered
moving vortex lattice.
This result is interesting since it indicates that
moving vortex phases do not always organize into ordered states. 
A time trace of a PML state
at fixed $F_{D}$ shows much weaker 
velocity oscillations than those shown in Fig.~4(c) for the ML state. 
This suggests that phase locking with PML states
will be very weak or absent. 
In previous work on phase locking for square arrays, it was shown that 
the phase locking is most pronounced 
at commensurate fields where the moving vortex structures are more ordered    
\cite{Rosseel2,Jensen2}.

\begin{figure}
\includegraphics[width=3.5in]{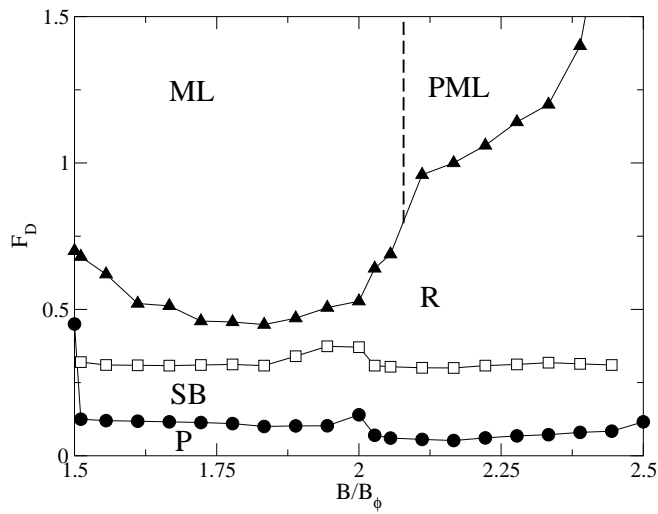}
\caption{
The dynamic phase diagram for $F_{D}$ vs $B/B_{\phi}$ highlighting 
the different dynamical phases. P: pinned, SB: symmetry broken, R: random,
ML: moving locked. Here $F_p=0.85$, $R_p=0.35\lambda$, and
$n_p=0.3125/\lambda^2$.
For $B/B_{\phi} > 2.125$, at high $F_{D}$ 
the system forms a partially moving locked (PML) phase where not all of the
vortices move along the pinning rows. 
}
\end{figure}

By performing a series of 
simulations for varied $B/B_{\phi}$, 
measuring the features in the velocity force curves and 
observing the vortex structures, 
we construct the dynamical phase diagram of  
$F_{D}$ vs $B/B_{\phi}$ shown in Fig.~10.
The depinning force marking the end of the pinned (P) phase
shows peaks at $B/B_{\phi} = 1.5$, 2.0, and $2.5$, corresponding to 
the commensurate and ordered ground states reported previously \cite{Pinning}. 
The SB-R transition line is fairly flat 
as a function of $B/B_\phi$ with an 
enhancement to higher values of $F_{D}$ occurring near $B/B_{\phi} = 2.0$, while
at the incommensurate fields,
monomers or trimers create fluctuations 
that cause the vortices at the pinning sites
to depin at slightly lower values of $F_{D}$. For $B/B_{\phi} < 2.1$, 
upon increasing $F_D$ the  
random state organizes into a ML state 
where all the vortices move along the pinning rows as shown 
in Fig.~3(e), 
while for $B/B_{\phi} \ge 2.1$, the random state organizes into the PML state.
The width of the random phase, as determined by the fluctuations in the 
velocity, increases and persists to higher values of
$F_{D}$ for increasing $B/B_\phi$ at $B/B_{\phi} \ge 2.1$.
For $B/B_{\phi} < 1.5$ and $B_{\phi} > 2.5$, where dimers are 
no longer present, the SB phase is lost
and a new set of dynamical phases 
arises which we discuss in more detail below.

\begin{figure}
\includegraphics[width=3.5in]{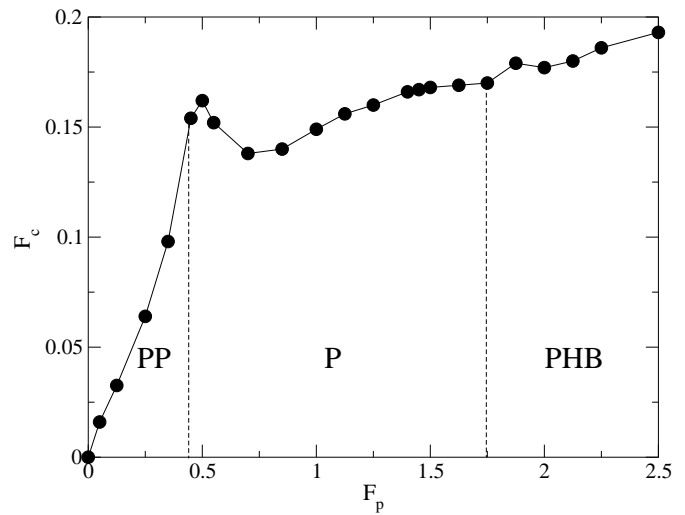}
\caption{
The critical depinning force $F_{c}$ vs $F_{p}$ 
for a system with $B/B_{\phi} = 2.0$, $R_p=0.35\lambda$, and
$n_p=0.3125/\lambda^2$.
For $F_{p} < 0.45$ a partially pinned (PP) state forms which is
illustrated in Fig.~12(a). 
For $0.45 \le F_{p} < 1.75$ the system forms 
the pinned (P) orientationally ordered
dimer state such as that shown in Fig.~1(c). 
For $F_{p} \ge 1.75$, the pinned herringbone (PHB) 
state seen in Fig.~12(b) forms.        
}
\end{figure}

\begin{figure}
\includegraphics[width=3.5in]{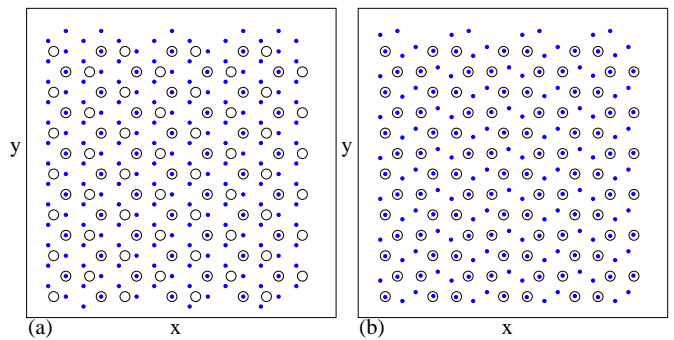}
\caption{
Vortex positions (filled circles) and 
pinning site locations (open circles) for the 
system in Fig.~11. 
(a) The partially pinned (PP) state at 
$F_{p} = 0.25$.  The vortex lattice structure consists of
a triangular lattice, and only half of the pinning sites are occupied. 
(b)  The pinned herringbone (PHB) state at $F_{p} = 2.0$. 
Here the dimers do not all align in the same direction, but instead
alternate in their alignment from row to row.
}
\end{figure}

\begin{figure}
\includegraphics[width=3.5in]{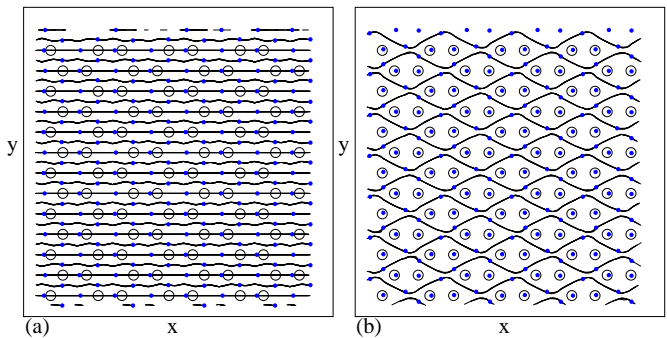}
\caption{
Vortex positions (filled circles), 
pinning site locations (open circles), and vortex trajectories for the
system in Fig. 11.
(a) Moving crystal (MC) state for $F_p=0.25$ and $F_{D} = 0.2$. 
The partially pinned (PP) state from Fig.~12(a) 
depins elastically into a MC
where half of the vortices move directly along the pinning rows
and the other half of the vortices move in winding paths between the
pinning sites. 
(b) Moving interstitial (MI) state for $F_p=2.0$ and $F_{D} = 0.225$. 
The pinned herringbone (PHB) state depins into 
a MI state in which interstitial vortices move around the pinned vortices.   
}
\end{figure}

\subsection{Effect of Changing the Pinning Strength} 

We next consider the effect of changing the pinning strength when  
$B/B_{\phi} = 2.0$.
The four phases in Fig.~2 occurred in a sample with 
$F_{p} = 0.85$.
As $F_{p}$ is varied, we find several different kinds of ordering 
within the pinned phase that
affect the dynamics which occur at finite $F_{D}$. 
In Fig.~11 we plot the threshold depinning force $F_{c}$ 
as a function of $F_{p}$. For $F_{p} < 0.35$ the pinning is weak enough 
that the vortex-vortex interactions dominate over the pinning energy
and a nearly triangular vortex lattice forms, as shown in Fig.~12(a).
In this arrangement, half of the pinning sites are still occupied, so the
vortex lattice is partially pinned (PP) and there is a finite depinning
threshold.
This type of partially pinned vortex lattice was observed in 
previous simulations
on honeycomb pinning lattices \cite{Pinning}, 
and  similar partially pinned vortex lattice states
have been predicted for square pinning arrays \cite{Peeters} 
and observed for metallic particles on 
periodic structures \cite{Coupier}. 
The depinning transition 
from the PP state is elastic, and 
all the vortices depin simultaneously to form
the moving triangular 
crystal (MC) shown in Fig.~13(a).
In the MC, half of the vortices move in one-dimensional paths along the pinning
rows while the remaining vortices 
move through the interstitial regions with a small transverse oscillation. 

Figure 11 shows that the pinned ordered dimer state (P) forms for 
$0.45 \leq F_p < 1.75$.
Over the range
$0.45 \leq F_p < 0.55$,
the depinning from state P does not occur by 
the initial flow of the interstitial vortices
into the SB phase, unlike the case shown earlier for $F_p=0.85$. 
Instead, for $0.45 \leq F_p < 0.55$,
both the interstitial vortices and the vortices at the pinning
sites depin simultaneously and rearrange into the moving crystal (MC) state
shown in Fig.~13(a).
We also find a peak in $F_{c}$ at $F_p=0.5$. 
This peak occurs due to both the change in the pinning configuration and a
change in the depinning process.  For $F_p<0.45$, only half of the pinning sites
are occupied and the vortex lattice depins elastically.
At $0.45 \leq F_p < 0.55$, all of the pinning sites are now occupied in the
P state, but the vortex lattice still depins elastically.  The pinning energy
that must be overcome to depin the lattice is increased compared to the
PP state, leading to an increase in $F_c$.  For $0.55 \le F_p < 1.75$, the
depinning process is plastic and only the interstitial vortices flow at
depinning to form the SB state.  Since the plastic depinning process does
not require pinned vortices to depin, the threshold force $F_c$ drops,
producing the peak in $F_c$ at $F_p=0.5$.

For $F_p\ge 0.7$ in Fig.~11, 
the depinning threshold $F_c$ slowly increases with
increasing $F_{p}$, and a transition in the pinned vortex structure 
occurs at $F_{p} = 1.75$. 
For $F_p \ge 1.75$, 
the dimers in the pinned state are no longer aligned 
but form a pinned herringbone (PHB) type structure such as that shown 
in Fig.~12(b), where
the dimers are tilted in the same direction 
in one row and tilted in the opposite direction in the adjacent 
rows. 
Herringbone ordering of dimers 
has previously been observed for colloidal dimers 
on triangular lattices \cite{Reichhardt} and for 
vortices in kagom{\' e} arrays at $5/3$ filling \cite{Pinning}. 
At depinning, the PHB state does not form a SB phase but instead
forms the winding interstitial phase shown in
Fig.~13(b). 
The dimers break apart into two monomers, with one 
monomer passing around the pinned vortices in the positive $y$ direction and
the other monomer passing the pinned vortices in the negative $y$ direction.
We term this state the moving interstitial (MI) phase.

In contrast to the herringbone state,
the aligned dimer or ferromagnetic ordering of the P state
occurs when the pinned vortices adjacent to the dimers are able to undergo
a periodic distortion within the pinning sites, reducing the interaction
energy between the pinned and interstitial dimer vortices and permitting
the dimer alignment.
If the dimers are aligned along $+30^{\circ}$, 
as in Fig.~5(a) of Ref.~\cite{Pinning}, the two pinned vortices 
closest to each interstitial vortex in the dimer both splay outwards away
from the $+30^{\circ}$ direction.
When $F_p$ is increased, the pinned vortices are pulled toward the center of
each pinning site and are no longer able to distort in order to accommodate
the aligned dimer state, so the pinned herringbone (PHB) state forms instead.
This result suggests that there may be other types 
of ground state ordering for 
vortices in honeycomb and kagom{\' e} arrays in addition to those that
have been reported previously.
It may be possible to use the size and shape of the pinning sites 
as a means of controlling the type
of crystalline structure that forms.

\begin{figure}
\includegraphics[width=3.5in]{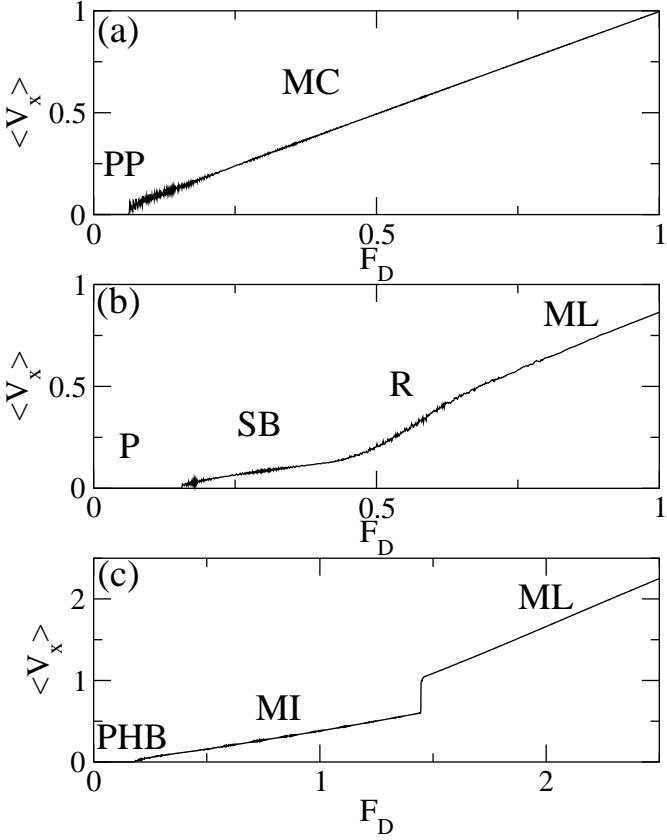}
\caption{
$\langle V_{x}\rangle$ vs $F_{D}$ 
for $B/B_{\phi} = 2.0$, $R_p=0.35\lambda$, and
$n_p=0.3125/\lambda^2$.  
(a) At $F_{p} = 0.25$, there is an elastic 
depinning transition between the     
partially pinned (PP) state and the moving crystal (MC) state. 
(b) At $F_{p} = 1.125$ the four dynamical phases are present.
P: pinned; SB: symmetry broken; R: random; ML: moving locked. 
(c) At $F_{p} = 2.125$ the pinned herringbone (PHB) state 
depins into the moving interstitial (MI) state
illustrated in Fig.~13(b). 
The transition between the MI and the moving locked (ML) state is 
much sharper than the R to ML transition shown in (b).
}
\end{figure}

In Fig.~14, we plot $\langle V_{x}\rangle$ 
versus $F_{D}$ for $F_{p} = 0.25$, 1.125, and
2.125. 
Figure 14(a) shows the elastic depinning process for the 
PP state at $F_p=0.25$, which moves directly into the MC phase
after depinning. 
At $F_{p} = 1.125$ in Fig.~14(b), 
$\langle V_x\rangle$ increases linearly with $F_D$ 
through the SB phase.  The slope
of $\langle V_{x}\rangle$ increases in the R phase, 
and the velocity-force relationship becomes linear in the ML phase.     
For $F_{p} = 2.125$, Fig.~14(c) indicates that the 
depinning occurs in two steps. 
The first depinning transition of the interstitial vortices only
takes the system from the P phase into the
moving interstitial (MI) phase, while at the second depinning transition,
the pinned vortices depin and the sample enters the ML phase.
Unlike the behavior at $F_p=1.125$ in Fig.~14(b), at $F_p=2.125$ 
the intermediate
random (R) phase
is lost and is replaced by a sharp jump into the ML phase. 

\begin{figure}
\includegraphics[width=3.5in]{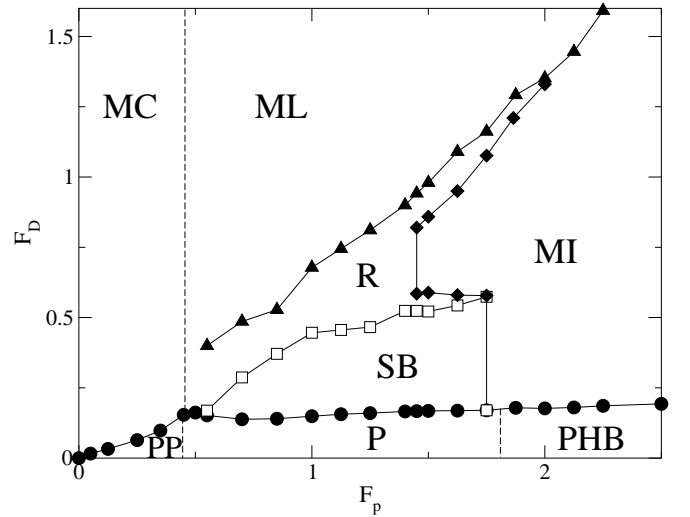}
\caption{
The dynamic phase diagram of $F_{D}$ vs $F_{p}$ for $B/B_{\phi} = 2.0$,
$R_p=0.35\lambda$, and $n_p=0.3125/\lambda^2$. 
The three pinned phases are: PP, 
the partially pinned phase shown in Fig.~12(a);
P, the pinned ordered dimer phase in Fig.~1(c); and PHB,
the pinned herringbone phase illustrated in Fig.~12(b). 
The dashed line separates the 
moving crystal phase (MC) shown in Fig.~13(a) 
from the moving locked (ML) phase of
Fig.~3(e). The symmetry broken (SB) phase 
illustrated in 
Fig.~3(a) occurs at intermediate vales of $F_{p}$, 
while the moving interstitial (MI) phase 
shown in Fig.~13(b) forms at higher values of $F_{p}$. 
The random (R) phase is illustrated in Fig.~3(c).       
}
\end{figure}

By conducting a series of simulations
we construct the dynamical phase diagram 
as a function of $F_p$ and $F_D$, as shown in Fig.~15. 
At high $F_D$, the MC phase forms for $F_p<0.45$, while for $F_p \ge 0.45$ the
ML phase appears instead.
The SB phase exists for
$0.55 < F_{p} < 1.75$, 
and the SB-R boundary shifts to higher $F_D$ with increasing $F_{p}$ until
it terminates at $F_{p} = 1.75$. 
For $F_p \ge 1.75$, the PHB state occurs at low drive, and the system depins
into the MI phase.
The MI phase also extends as far down as $F_p=1.5$, where the system
passes from the SB phase into a narrow window of the random (R) phase
with increasing $F_D$ before
the vortices organize into the MI phase. 
As $F_{D}$ continues to increase,
the vortices at the pinning sites depin 
and the system passes through a second narrow window of the R phase until
the vortices 
organize into the ML phase. 
At high $F_{p}$, the R phase becomes
vanishingly small and the system passes directly 
from the MI to the ML phase. 
The transition into the ML state increases linearly with 
increasing $F_{p}$, while the
depinning force saturates with increasing $F_{p}$.      

\begin{figure}
\includegraphics[width=3.5in]{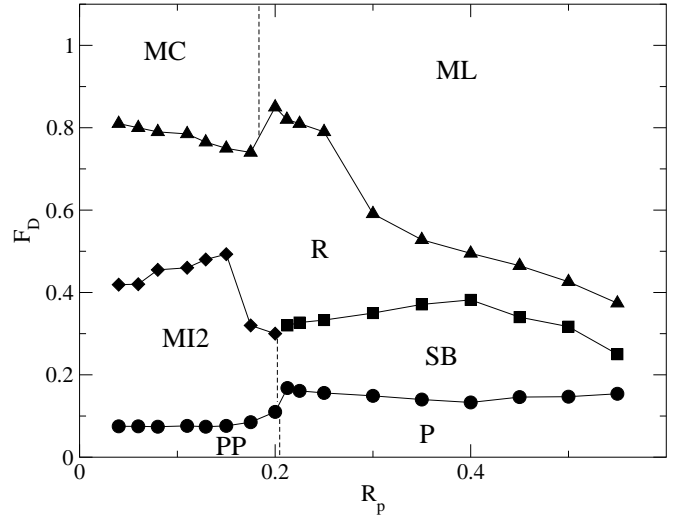}
\caption{
The dynamic phase diagram of $F_{D}$ vs $R_{p}$ for  $B/B_{\phi} = 2.0$ 
and $F_{p} = 0.85$.
PP: partially pinned phase; P: pinned dimer phase; MI2: moving interstitial
phase 2; SB: symmetry broken phase; R: random phase; MC: moving crystal phase;
ML: moving locked phase.
For $R_{p} > 0.2\lambda$ we observe
the same phases illustrated in Fig.~15
at $F_p=0.85$. 
The curves do not extend above $R_p=0.55\lambda$ since for $R_p>0.55\lambda$,
multiple vortex pinning at individual pinning sites occurs. 
For $R_{p} < 0.2\lambda$, 
a partially pinned phase appears 
and the initial depinning is into a new 
moving interstitial phase termed MI2, 
illustrated in Fig.~17(a). The upper dashed 
line separates the MC phase from the ML phase.    
}
\end{figure}

\begin{figure}
\includegraphics[width=3.5in]{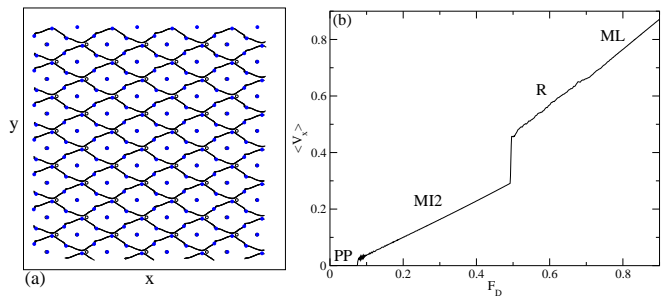}
\caption{
(a) Vortex positions (filled circles), pinning site locations 
(open circles)  and vortex trajectories (black lines) for the system
in Fig.~16 showing the moving
interstitial 2 phase (MI2) at $R_{p}= 0.15\lambda$ and $F_{p} = 0.2$. 
(b) $\langle V_{x}\rangle$ vs $F_{D}$ for the same system.
A sharp transition from the MI2 phase to the random (R) phase occurs.   
}
\end{figure}

\subsection{Changing $R_{p}$ and $B_{\phi}$} 

We next examine the effects of 
changing the pinning radius in a system with fixed $F_{p} = 0.85$ and
$B/B_{\phi} = 2.0$. 
In Fig.~16 we show the dynamic phase diagram for $F_{D}$ versus $R_{p}$ 
obtained from a series of simulations.
For $0.2\lambda < R_{p} \le 0.55\lambda$, 
we find the same three moving phases, SB, R, and ML, as in Fig.~2. 
For $R_{p} > 0.55\lambda$, the pins are large enough to permit 
double vortex occupancy at the individual pinning sites; in this case, 
a new set of phases appears which we do
not consider in this work.
For $R_{p} > 0.2\lambda$, the R-ML transition decreases in $F_D$ with 
increasing $R_p$ since
the larger pinning sites make it easier 
for the vortices to localize along the    
pinning rows and flow in the one-dimensional motion of the ML phase. 
For $R_{p} < 0.2\lambda$ at low $F_D$, 
we find the partially pinned PP state illustrated
in Fig.~12(a). The onset of the PP phase
coincides with a drop in $F_{c}$ at $R_{p} \approx 0.2\lambda$. 
Unlike the PP phase that occurs 
at low $F_p$ in Fig.~15, which depins elastically,
the PP phase at small $R_p$
depins into a moving interstitial phase that is distinct from the
moving interstitial phase shown in Fig.~13(b). 
In Fig.~17(a) we illustrate 
the vortex trajectories in the phase which we term 
the moving interstitial phase 2 (MI2). 
The vortices flow in winding interstitial channels; however, unlike
the MI phase, in the MI2 phase only half of the pinning sites are occupied.
In Fig.~17(b) we plot $\langle V_{x}\rangle$ vs $F_{D}$ for a system
with $R_{p} = 0.15\lambda$.  A clear two-step depinning transition 
occurs, with $\langle V_{x}\rangle$ increasing
linearly with increasing $F_D$ in the MI2 phase. 
At high $F_D$ and $R_p<0.2\lambda$, 
the pinning sites are too small for the ML phase to occur, and instead 
the vortices flow in the MC phase illustrated in Fig.~13(a).

\begin{figure}
\includegraphics[width=3.5in]{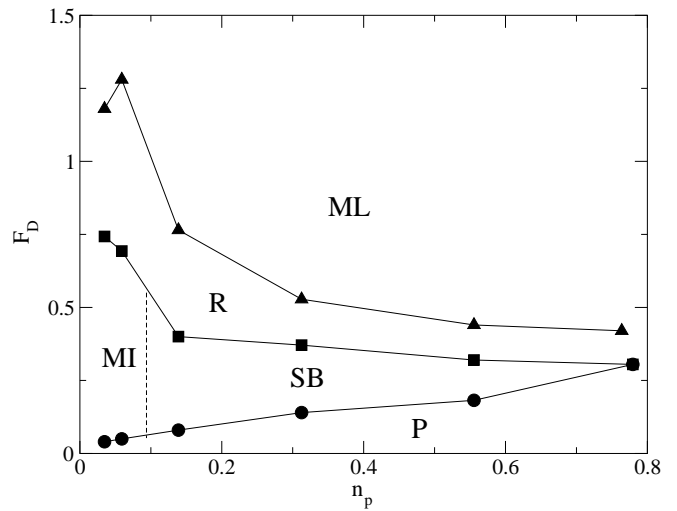}
\caption{
The dynamic phase diagram $F_{D}$ versus 
pinning density $n_v$, which determines $B_{\phi}$, 
at $B/B_{\phi} = 2.0$, $F_{p} = 0.85$,
and $R_{p} = 0.35\lambda$. 
P: pinned dimer phase; MI: moving interstitial phase; SB: symmetry broken 
phase; R: random phase; ML: moving locked phase.
The transition to the ML phase shifts to higher 
$F_{D}$ with decreasing $n_p$.
The SB phase appears only for intermediate values
of 
$n_p$.
}
\end{figure}

We next consider samples with fixed $B/B_{\phi} = 2.0$, 
$R_{p} = 0.35\lambda$, and $F_{p} = 0.85$, but 
vary the value of $B_{\phi}$ by changing the pinning density $n_p$.
This alters the average spacing between neighboring vortices.
Up to this point we have used $n_p=0.3125/\lambda^2$.
In Fig.~18 we illustrate
the dynamic phase diagram for 
$F_D$ versus $n_p$.
As 
$n_p$ increases, $F_{c}$ increases 
since the depinning of the interstitial vortices is determined by 
the potential created by the vortices located at the pinning sites, 
and as the vortex density increases, the
depth of the interstitial pinning potential 
also increases. 
The R-ML transition shifts to higher $F_D$ with decreasing $n_p$.
Since the distance between the pinning sites increases with
decreasing 
$n_p$, the moving vortices 
spend less time in the pinning sites. 
This destabilizes the ML phase
and the vortices must move at higher velocities
for the effective trough potential to be able to stabilize
the ML phase. 
For 
$n_p \geq 0.78/\lambda^2$, the SB phase is lost, since at this 
pinning density the interactions between the interstitial vortices 
and the pinned vortices become sufficiently strong
that the depinning of the interstitial vortices also causes the pinned vortices
to depin.  As a result, the system passes directly from the P phase to the
R phase with increasing $F_D$.
For 
$n_p<0.14/\lambda^2$,
the vortex-vortex interaction becomes weak enough that the system depins
into the MI phase illustrated in Fig.~13(b).
This also coincides with an 
increase in the value of $F_{D}$ at which the R phase appears,
since the moving interstitial vortices in the MI phase 
do not approach the pinned vortices
as closely as they do in the SB phase.  

\begin{figure}
\includegraphics[width=3.5in]{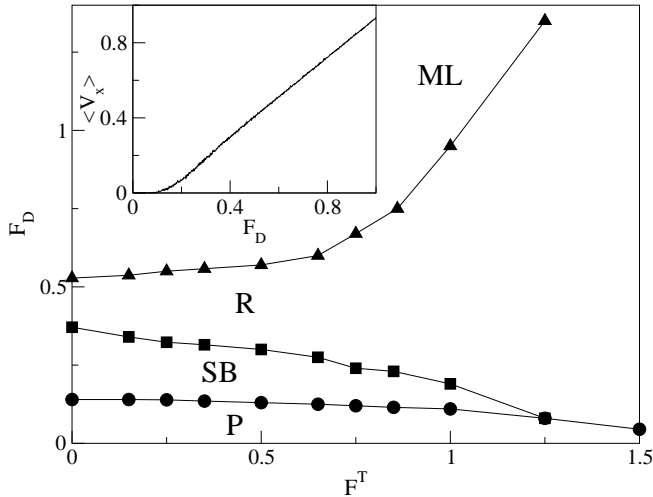}
\caption{
The dynamic phase diagram of $F_{D}$ vs 
thermal force $F^{T}$ for a system with $B/B_{\phi} = 2.0$, $F_{p} = 0.85$,
$n_p=0.3125/\lambda^2$,
and $R_{p} = 0.35\lambda$.
P: pinned dimer phase; SB: symmetry broken phase; R: random phase; and ML:
moving locked phase. 
Inset: $\langle V_{x}\rangle$ vs $F_{D}$ at $F^{T} = 1.25$ where a smooth
depinning transition occurs.
}
\end{figure}

\begin{figure}
\includegraphics[width=3.5in]{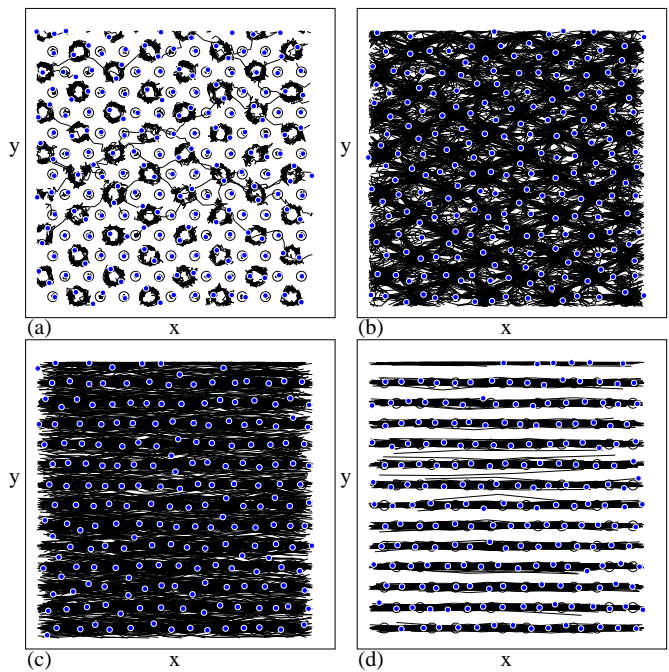}
\caption{
Vortex positions (filled circles), pinning site locations (open circles), 
and vortex trajectories (black lines)
for the system in Fig.~19 at $F^{T} = 1.25$. 
(a) At $F_{D} = 0.1$,  there 
is appreciable creep of the interstitial vortices and
the dimers have lost their orientational ordering and are rotating within the
large interstitial sites.
(b) At $F_{D} = 0.2$, the vortex flow is disordered 
and vortices are continually depinning and repinning. 
(c) At $F_{D} = 1.0$, all of the vortices are moving but there is still
diffusion in the direction transverse to the drive.
(c) At $F_{D} = 1.6$ the vortices are about to enter the ML state
where the motion is confined to one-dimensional channels.
}
\end{figure}

\subsection{Effect of Finite Temperature}

We next 
consider the effect of finite temperature 
on the system in Fig.~2 with $B/B_\phi=2.0$, $F_p=0.85$, $R_p=0.35\lambda$,
and $n_p=0.3125/\lambda^2$.
In Ref.~\cite{Pinning}, we showed that a transition can occur 
at finite temperature in which the
vortex $n$-mer states lose their orientational ordering and 
begin to rotate while remaining confined within the 
large interstitial sites. 
This state was termed a vortex plastic crystal. 
In Ref.~\cite{Transverse} we demonstrated that the SB phase
disappears in the vortex plastic crystal state.
In Fig.~19 we plot the dynamical phase diagram 
of $F_{D}$ vs $F^{T}$ for the same system in Fig.~2. 
In our units, the dimers melt at $F^{T} = 1.0$. Above the melting
temperature, there  
is appreciable creep of the interstitial vortices as they hop
from one large interstitial site to another, 
as illustrated in Fig.~20(a) for $F_{D} = 0.1$ and $F^{T} = 1.25$.
As $F_{D}$ is further increased
the system enters the random (R) phase shown in Fig.~20(b) 
for $F_{D} = 0.25$.
At higher drives, the vortices begin to localize along the pinning rows
in one-dimensional channels; however, 
there is still appreciable hopping from one row to another as 
shown in Fig.~20(c) for $F_{D} = 1.0$. 
At even higher drives,
the ML is recovered as illustrated in Fig.~20(d) 
for $F_{D} = 1.6$. 
For $F^{T} > 1.35$, the ML phase is lost and the
high $F_{D}$ flow is in the random (R) phase
shown in Fig.~20(c). 
In the inset to Fig.~20 we demonstrate that
at $F^{T} = 1.25$, the sharp features in the velocity force curve 
seen at $F^{T} = 0$ in 
Fig.~2 disappear.

\begin{figure}
\includegraphics[width=3.5in]{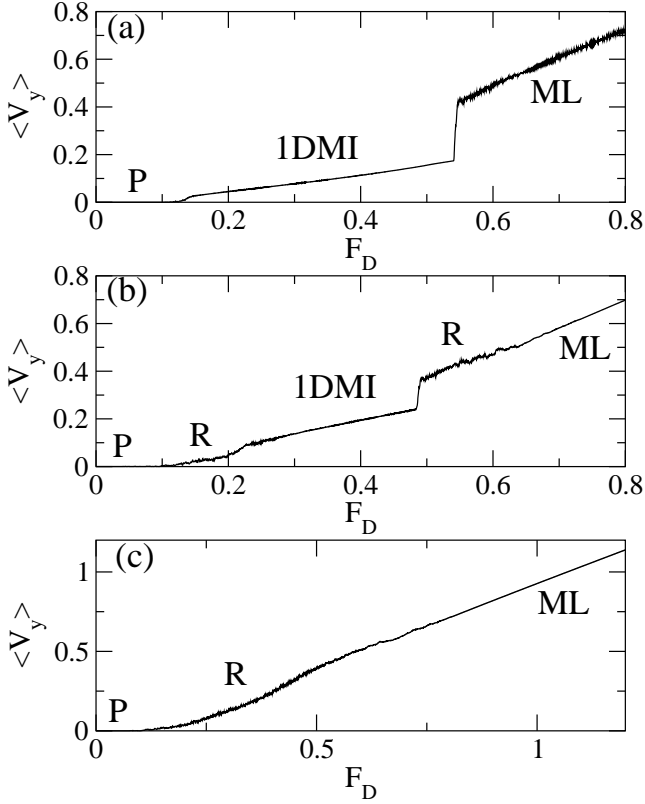}
\caption{
The velocity force curves 
$\langle V_y\rangle$ vs $F_D$
for ${\bf F}_D=F_D{\bf {\hat y}}$
for the system in Fig.~2
with $F_{p} = 0.85$, $R_{p} = 0.35\lambda$,
and $n_p=0.3125/\lambda^2$. 
P: pinned dimer phase; R: random phase; 1DMI: one-dimensional moving 
interstitial phase, illustrated in Fig.~22(a); ML: moving locked phase,
illustrated in Fig.~22(b).
(a) $B/B_{\phi} = 2.0$.
(b) $B/B_{\phi}= 2.15$, where 
an additional random flow phase occurs at 
depinning due to the presence of trimers.  
(c) $B/B_{\phi} = 2.32$, where there is no longer a 1DMI phase.
}
\end{figure}

\begin{figure}
\includegraphics[width=3.5in]{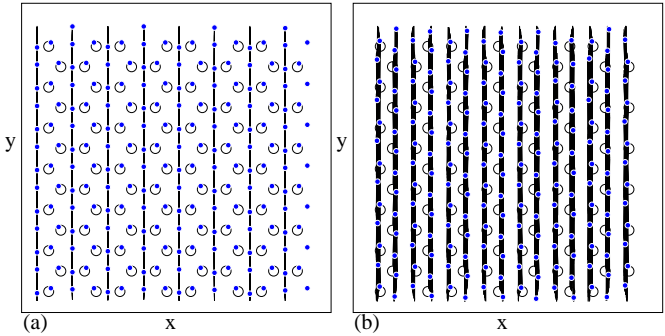}
\caption{
Vortex positions (filled circles), 
pinning site positions (open circles), and vortex trajectories (black lines)
for the system in Fig.~21(a) at $B/B_\phi=2.0$. 
(a) One-dimensional moving interstitial (1DMI) phase at $F_{D} = 0.2$. 
(b) The moving locked (ML) phase for 
${\bf F}_D=F_D{\bf {\hat y}}$ 
at $F_{D} = 0.8$.    
}
\end{figure}

\section{Dynamics for Driving in the Transverse Direction}

We now consider the case where $F_{D}$ is applied along the $y$-direction, 
${\bf F}_D=F_D{\bf {\hat y}}$, for the same system as in 
Fig.~1(d) with $F_p=0.85$, $R_p=0.35\lambda$, and $n_p=0.3125/\lambda^2$. 
A different set of dynamic phases appear that are
distinct from those found for driving in the $x$-direction. 
In particular, 
the SB phase is lost and the dimers align in the $y$-direction with
the initial depinning occurring in one-dimensional 
interstitial flow paths. 
In Fig.~21 we plot the velocity force curves for $B/B_{\phi}=2.0,$
2.15, and 2.32.

Figure 21(a) shows the three phases, pinned (P), one-dimensional moving
interstitial (1DMI), and moving locked (ML) that occur at
$B/B_{\phi} = 2.0$. 
In the P phase, if the ground state contains dimers which are aligned at 
either $+30^\circ$ or $-30^\circ$ to the $x$-axis,
a polarization effect is induced by the applied drive similar
to the effect discussed earlier.  In this case, however, the dimers shift 
such that they are aligned in the
$y$-direction. 
The one-dimensional moving interstitial (1DMI) 
state which appears above depinning is
illustrated in Fig.~22(a) at $F_D=0.2$, where 
the interstitial vortices move 
between the vortices in the pinning sites. 
Near $F_{D} =0.55$ there is a sharp depinning 
transition for the vortices in the pinning sites.  
After this depinning transition occurs, the vortices very rapidly
rearrange into a moving locked (ML) phase where the vortices move along 
the pinning sites, as shown in Fig.~22(b) for $F_{D} = 0.8$. 
In the ML phases for driving along the $x$ and the $y$ directions,
the vortices travel in one-dimensional channels along a row or a column
of pinning sites, respectively.
The pinning rows followed by the vortices for $x$-direction driving are
evenly spaced in the $y$-direction, so the vortices flow through the
centers of the pinning sites.  In contrast, the pinning columns
followed by the vortices for $y$-direction driving are unevenly spaced
in the $x$ direction due to the symmetry of the honeycomb lattice.
As a result, for the $y$-direction driving the vortices do not flow through
the centers of the pinning sites, but are instead shifted to the right and
left of the pinning sites in alternate columns, as seen in Fig.~22(b).  
This produces a more
even spacing between the columns of moving vortices.
Since different columns contain different
numbers of vortices, the ML phase for $y$-direction driving 
has smectic type characteristics.
The velocity-force curves
for $1.5 < B/B_{\phi} < 2.0$ 
have the same general form as the curve in
Fig.~21(a) and show the same three phases.

\begin{figure}
\includegraphics[width=3.5in]{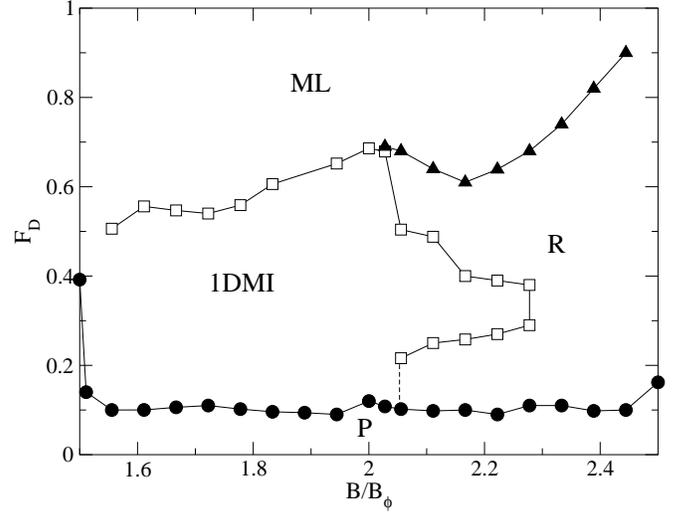}
\caption{
The dynamical phase diagram for $F_{D}$ vs $B/B_{\phi}$ for 
${\bf F}_D=F_D{\bf {\hat y}}$ in a sample
with $F_{p} = 0.85$, $R_{p} = 0.35\lambda$, and
$n_p=0.3125/\lambda^2$. 
P: pinned phase; 1DMI: one-dimensional moving interstitial phase;
R: random phase; ML: moving locked phase.
For $B/B_{\phi} > 2.3$ the 1DMI flow is lost.  
}
\end{figure}

For $B/B_{\phi} > 2.0$ the appearance of trimer states disrupts the 
1DMI flow since the trimers cannot align completely in the
$y$-direction. 
This produces random (R) vortex flow at depinning, as shown in Fig.~21(b),
with diffusive vortex motion occurring along the $x$-direction. 
There are more pronounced fluctuations in $\langle V_y\rangle$ in the
R phase, and the velocity-force curve is nonlinear and lower than the 
extrapolated linear behavior in the 1DMI phase that begins near 
$F_D=0.23$.
The trimers can block the one-dimensional
channels of flow shown in Fig.~22(a), lowering the 
number of mobile vortices. 
At higher drives, the trimers depin, straighten
into a linear configuration, and flow in the 1DMI phase.
At $F_D \approx 0.5$,
the vortices in the pinning sites depin, resulting in a transition 
from the 1DMI phase to the R phase.
For sufficiently high drives, the ML phase forms.
For driving along the $x$-axis at $B/B_\phi>2.0$, 
we showed in Fig.~9(c) that
the ML phase is lost due to a buckling transition of the one-dimensional
chains of vortex motions, and that a partially moving locked (PML) phase
forms instead when a portion of the vortices move through the interstitial
regions.
For driving along the $y$-axis at $B/B_\phi>2.0$, 
the ML state remains stable for
much higher values of $B/B_{\phi}$ than for the $x$-axis case.
A comparison of Fig.~22(b) and Fig.~3(e) 
shows that the interstitial region crossed by the vortices
in the moving channels is not as wide for $y$-axis driving as 
for the $x$-axis driving, resulting in more stable $y$-axis ML flow.
As $B/B_{\phi}$ is further increased, the 
random regime grows until the 
1DMI phase is completely lost, as shown in Fig.~21(c)
for $B/B_{\phi} = 2.35$. 
In Fig.~23 we plot the
dynamical phase diagram for 
$F_{D}$ versus $B/B_{\phi}$, 
highlighting the onset of the different phases. 
The transition to the ML phase 
shifts to higher values of $F_D$ with increasing $B/B_\phi$ since
the ML vortex channels become increasingly anisotropic as the number of 
vortices in the sample increases.

\begin{figure}
\includegraphics[width=3.5in]{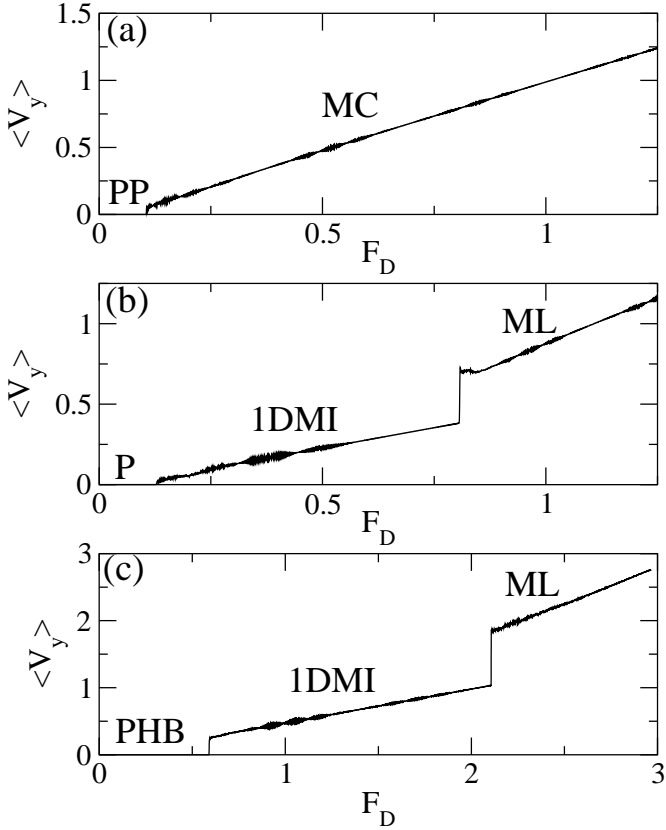}
\caption{
$\langle V_{y}\rangle$ 
vs $F_{D}$ for 
${\bf F}_D=F_D{\bf {\hat y}}$ for
$B/B_{\phi} = 2.0$, $R_{p} = 0.35\lambda$, and $n_p=0.3125/\lambda^2$.  
(a) At $F_{p} = 0.35$, there is a single step elastic depinning transition
from the partially pinned (PP) phase to the moving crystal (MC) phase.
(b) At $F_{p} = 1.25$ we find the pinned (P), one-dimensional moving
interstitial (1DMI), and moving lattice (ML) phases.
(c) At $F_{p} = 2.25$, there are sharp transitions between the
pinned herringbone (PHB), one-dimensional moving interstitial (1DMI), 
and moving locked (ML) phases.
}
\end{figure}

\begin{figure}
\includegraphics[width=3.5in]{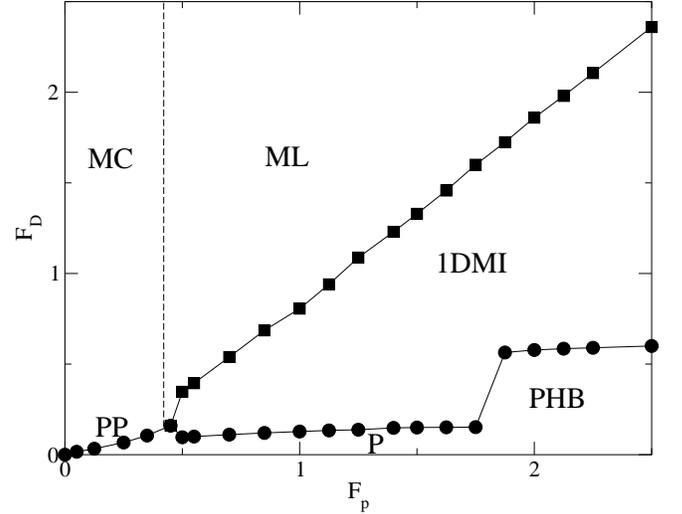}
\caption{
The dynamic phase diagram 
$F_{D}$ vs $F_{p}$ for 
${\bf F}_D=F_D{\bf {\hat y}}$
at $B/B_{\phi} = 2.0$, $R_{p} = 0.35\lambda$, 
and $n_p=0.3125/\lambda^2$.
PP: partially pinned phase; P: pinned dimer phase; PHB: pinned herringbone
phase; 1DMI: one-dimensional moving interstitial phase; MC: moving crystal
phase; ML: moving locked phase.
A peak in the depinning threshold $F_{c}$ 
occurs near $F_{p} = 0.5$ 
at the PP-P transition.
At the P-PHB transition,
$F_{c}$ 
increases by a factor of three. The dashed line separates the MC
phase from the ML phase.   
}
\end{figure}

\subsection{Dynamics as a function of $F_{p}$ and Dimer Jamming}  

We now consider the vortex dynamics in a system with
fixed $B/B_{\phi} = 2.0$, $R_p=0.35\lambda$, and
$n_p=0.3125/\lambda^2$ for varying $F_{p}$ with ${\bf F}_D=F_D{\bf {\hat y}}$. 
As noted above,
for $F_{p} < 0.5$ a partially pinned (PP) vortex lattice forms.
In Fig.~24(a), the velocity-force 
curve for $F_{p} = 0.35$ 
shows that the depinning of the PP phase is elastic and occurs in a
single step transition to a moving crystal (MC) phase,
as seen earlier in Fig.~14(a) for driving along the $x$-axis.  
In the MC phase, half of the vortices move along the pinning sites. 
For $0.5 < F_{p} < 1.75$, at low drive the system is in the pinned (P) phase
of orientationally ordered dimers, and as the drive increases, 
Fig.~24(b) shows that the same
1DM1 and ML phases illustrated in Fig.~21(a) appear.
The rapid rearrangement of the vortices 
from the 1DMI phase to the ML phase results in
a small jump near $F_{D}=0.8$ which
marks the 1DMI-ML transition.
For strong pinning $F_p \ge 1.75$, Fig.~24(c) indicates that
the ground state forms the pinned herringbone (PHB) phase 
illustrated in Fig.~12(b).  The same pinned state appears for $x$-direction
driving at strong pinning, as shown in Fig.~14(c).
In Fig.~24(c), the velocity-force curve 
at $F_{p} = 2.25$ shows the abrupt nature of the
depinning transition from the PHB phase to 
the 1DMI phase, which differs
from the smoother depinning transition 
that occurs from the P phase to the 1DMI phase in Fig.~24(b). 
The depinning threshold increases markedly with increasing $F_p$ once
the system enters the PHB state.
In Fig.~25 we plot
the dynamic phase diagram for $F_{d}$ versus $F_{p}$. 
Near the transition from the PP to the P 
phase, there is a peak in $F_{c}$ similar to 
the peak observed at the PP-P transition for driving in the $x$-direction
in Fig.~11. For $F_{p} > 0.175$ the  strong enhancement of the depinning  
threshold in the PHB state can be seen clearly.

\begin{figure}
\includegraphics[width=3.5in]{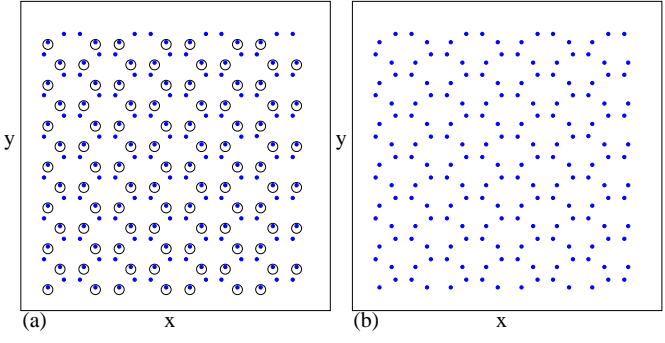}
\caption{
Vortex positions (filled circles) 
and pinning site locations (open circles) for the 
system in Fig.~24(c) at $B/B_{\phi} = 2.0$ and  $F_{D} = 0.56$, 
just before depinning. Under the influence of the driving force
which is applied in the $y$ direction,
the dimers align with the $x$ direction and shift to the top of the
large interstitial sites.
Because the
dimers are not aligned with the direction of the drive, 
a jamming phenomenon occurs which is responsible 
for the large increase in $F_{c}$ seen in Fig.~25 at $F_{p} = 1.75$. 
We call this the jammed 
state J. In (b) only the vortex positions are shown and it can more clearly
be seen that the dimers are shifted in the positive 
$y$-direction. The vortex configuration in the
jammed state is distinct from the pinned herringbone state. 
}
\end{figure}

In Fig.~26(a) we illustrate 
the vortex positions just before depinning for 
$F_{p} =  2.25$. Even though the drive is applied in the $y$-direction, the
dimers have aligned with the $x$-direction.
When the dimers are oriented along the $x$-axis,
they cannot fit through the easy-flow one-dimensional channel between 
the pinning sites,
but instead are essentially jammed by the two pinned vortices at the
top edge of the large interstitial site.
In the ordered dimer pinned (P) phase, the
dimers all reorient in the same direction under an applied drive.
In contrast, in the pinned herringbone (PHB) phase,
the dimers rotate in opposite directions 
under an applied drive, so 
when the drive is applied along the $y$-direction the 
dimers end up aligning in the $x$-direction. 
In Fig.~26(b) only the vortex positions from 
Fig.~26(a) are shown to indicate more clearly the shift of the dimers
in the positive $y$-direction.
This vortex configuration, which we term the jammed (J) state, has
a structure that is distinct from 
that of the pinned herringbone phase shown in Fig.~12(b). The 
jammed state configuration exists 
only in the presence of the applied drive. For $F_{d} = 0$ the 
dimers return to the pinned herringbone (PHB) state.         
In the jammed (J) state, 
the critical current is up to three times larger than in the state
where the dimers are aligned in the $y$-direction.
We also note that at incommensurate fields for $F_{p} > 1.75$, 
the net vortex flow is reduced 
since some of the dimers align in the $x$-direction and 
effectively block the 
motion of other vortices along the $y$-direction. 

\begin{figure}
\includegraphics[width=3.5in]{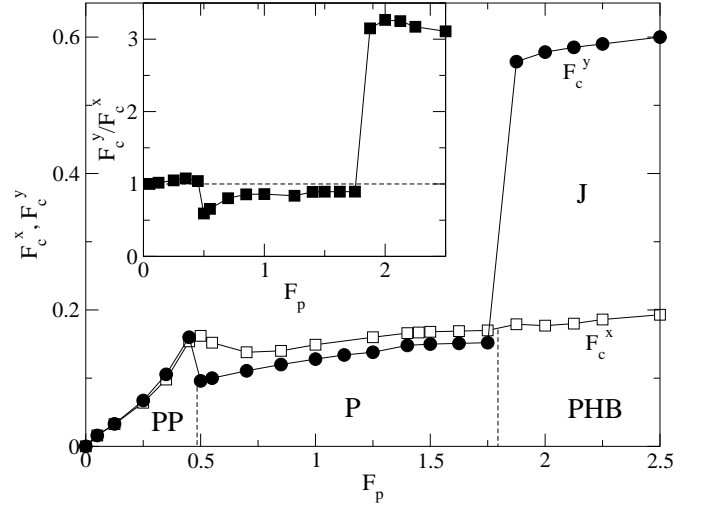}
\caption{
The critical depinning force in the $x$-direction, $F^{x}_{c}$ 
(open squares),
and in the $y$-direction, $F^{y}_{c}$ (filled circles), vs $F_{p}$
for $B/B_{\phi} = 2.0$, $R_{p} = 0.35\lambda$,
and $n_p=0.3125/\lambda^2$.  
PP: partially pinned phase; P: pinned dimer phase; PHB: pinned herringbone
phase; J: jammed state.
In the PP phase, $F_c^x=F_c^y$, while in the P phase, $F_c^x>F_c^y$.
A large enhancement of $F_c^y$ occurs in the PHB phase when dimer jamming
occurs.
Inset: the ratio $F^{y}_{c}/F^{x}_{c}$ vs $F_{p}$. The dashed line 
indicates $F^y_c/F^x_c=1$, where
the depinning thresholds are equal.
}
\end{figure}

In order to better characterize the enhancement of $F_{c}$ in the jammed state, 
in Fig.~27 we plot
the critical depinning force in the $y$-direction, $F^{y}_{c}$, and 
in the $x$-direction, $F^{x}_{c}$, versus $F_{P}$. 
In the inset of Fig.~27 we show the ratio $F^{y}_{c}/F^{x}_{c}$ versus $F_{p}$. 
In the partially pinned (PP) phase, 
$F_c^y=F_c^x$,
while in the pinned aligned dimer (P) phase,
$F_c^x$ is slightly higher than $F_c^y$ 
since the vortices can depin 
more readily into the 1DMI phase in the $y$-direction. 
In the jammed state that forms from the PHB phase,
$F^{y}_{c}$ is 3.1 times higher than $F^{x}_{c}$ for the same value of $F_{p}$. 

\begin{figure}
\includegraphics[width=3.5in]{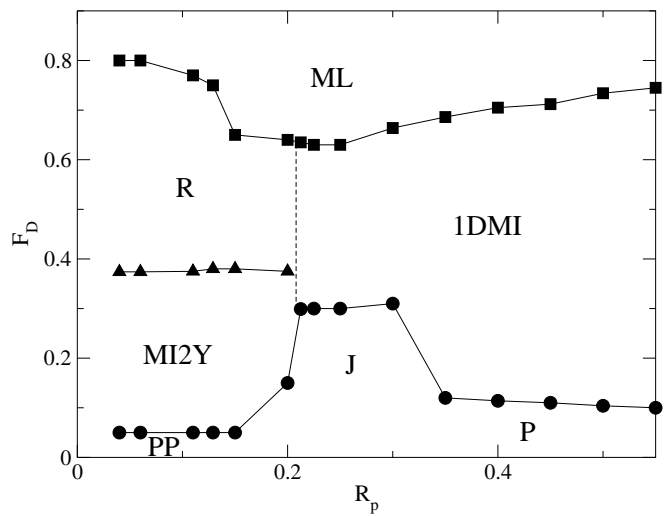}
\caption{
The dynamic phase diagram for $F_{D}$ vs $R_{p}$ 
with ${\bf F}_D=F_D{\bf {\hat y}}$, 
$B/B_{\phi} = 2.0$, $F_{p} = 0.85$, and $n_p=0.3125/\lambda^2$. 
PP: partially pinned phase; J: jammed state; P: pinned phase;
MI2Y: $y$-direction moving interstitial phase 2; 1DMI: one-dimensional
moving interstitial phase; R: random phase; ML: moving locked phase.
For $R_{p} < 0.2\lambda$ the system forms the PP
phase. This depins into the MI2Y state for driving in the $y$-direction, 
which is similar to the MI2 state shown in Fig.~17(a) 
for driving in the $x$-direction. 
For $0.2\lambda < R_{p} < 0.3\lambda$, the system forms
the jammed J state shown in Fig.~26.      
}
\end{figure}

\subsection{Effects of Changing $R_{p}$ and $B_{\phi}$}

In Fig.~28 we plot the dynamical phase diagram $F_D$ versus $R_p$ for 
driving in the $y$-direction 
with $B/B_{\phi} = 2.0$, $F_{p} = 0.85$, and $n_p=0.3125/\lambda^2$. 
For $R_{p} \ge 0.2\lambda$, 
the system depins into the 1DMI phase and makes a transition to 
the ML phase at higher drives. 
For $R_{p} < 0.2\lambda$, 
the system forms the partially pinned (PP) phase 
where only half of the pinning sites are occupied.
The PP phase depins into a moving interstitial 
phase (MI2Y) that resembles the MI2 state 
observed for driving in the $x$-direction in Fig.~17(b), 
where half the vortices depin while the other 
half remain pinned. The MI2Y phase is oriented $90^\circ$ from the MI2
phase.
At $F_{D} = 0.4$ for $R_p < 0.2\lambda$,
the vortices at the pinning sites begin to
depin and repin, giving a regime of the random (R) phase until 
$F_{D}$ becomes large enough for all the vortices to depin
into the ML phase. 
For $ 0.2\lambda \leq R_{p} \leq 0.3\lambda$, the jammed (J) state 
discussed in Fig.~26 occurs due to the formation of
dimers aligned in the 
$x$-direction, which is associated with a marked increase in $F_{c}$.
As discussed earlier, the pinned herringbone (PHB) phase and 
jammed (J) state occur when $F_p$ becomes high enough that the
vortices in the pinning sites cannot shift
to allow for dimer ordering to occur. 
Similarly, as $R_{p}$ is reduced, the vortices in the pinning sites 
have less room to adjust for dimer ordering, 
so the PHB state forms. 
The jamming also produces the counterintuitive effect 
that as $R_{p}$ increases above $R_p=0.35\lambda$, the 
depinning threshold decreases.

\begin{figure}
\includegraphics[width=3.5in]{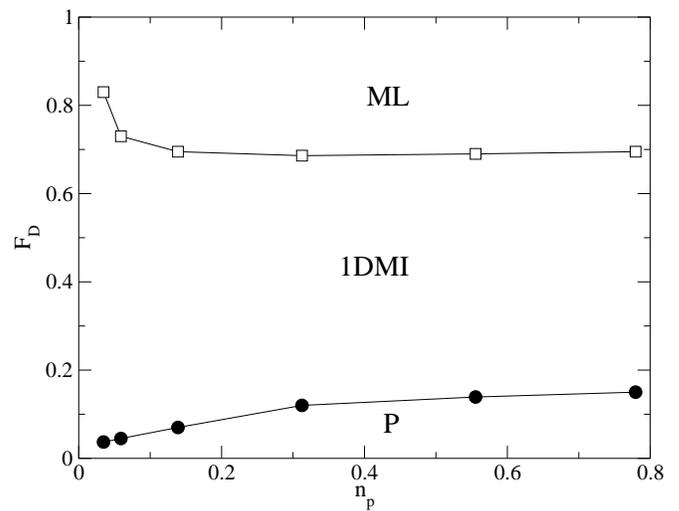}
\caption{
The dynamical phase diagram for $F_{D}$ 
vs $n_{p}$, which determines $B_\phi$, 
for driving in the $y$-direction with
$B/B_{\phi} = 2.0$, $R_{p} = 0.85\lambda$, and $F_{p} = 0.85$.
P: pinned ordered dimer phase; 1DMI: one-dimensional moving interstitial
phase; ML: moving locked phase.     
}
\end{figure}

In Fig.~29 we show 
the dynamical phase diagram for 
$F_{D}$ versus $n_p$, which determines the value of $B_{\phi}$, 
for $B/B_{\phi} = 2.0$, $R_{p} = 0.35\lambda$, and $F_{p} = 0.85$.  
As $n_p$ increases, 
the critical depinning force into the 1DMI phase 
increases since the repulsion from the pinned vortices experienced by the
interstitial vortices increases as the average vortex-vortex spacing decreases.
The transition from the 1DMI phase to the ML phase 
shifts to higher values of $F_D$ 
as $n_{p}$ decreases since the
distance between the pinning sites which stabilize the ML flow increases.

\section{Discussion}

Our results are for honeycomb pinning arrays 
where it was shown in previous
work that $n$-merization of the 
interstitial vortices into vortex molecular crystal states occurs for
$B/B_{\phi} > 1.5$.
Many of the dynamical effects presented in this work are due to 
the $n$-merization effect. 
In kagom{\' e} pinning arrays, 
similar types of vortex molecular crystal states
appear, so we expect that many of the same types of dynamic
phases described here will also occur for kagom{\' e} pinning arrays,
although we do expect that there will be 
certain differences as well. 
In the kagom{\' e} pinning array, the vortex dimer state 
appears at $B/B_{\phi} = 1.5$ and has 
a herringbone ordering even for large, weak pins. 
There are no easy flow channels along $\pm 30^\circ$ to the $x$-axis,
so the symmetry breaking flows should be absent.
Additionally, since there is no easy flow channel in the
$y$-direction, the
anisotropic depinning dynamics may be different as well.

We have only considered $B/B_{\phi} < 2.5$ in this work. 
At higher fields, 
a wide array of vortex molecular crystal states occur that should also have 
interesting dynamical phases. 
Since the low matching fields are more robust, observing the dynamics 
near these low fields experimentally is more feasible.
Although our results are 
specifically for pinning sites with single vortex occupation, 
similar dynamics should occur if the 
first few matching fields have multiple vortices at the pinning sites.
In this case, the effective dimerization of the interstitial
vortices would be shifted to higher magnetic fields.    

Although true phase transitions are associated only with 
equilibrium phenomena, the nonequilibrium 
phases considered here have many analogies to equilibrium phases. 
For example, several of the transitions
between the nonequilibrium phases have a continuous type 
behavior, while in other cases the transitions are sharp,
indicative of a first order nature. 
Future studies could explore the possible emergence of  
a growing correlation length near the 
transitions to see whether they exhibit the true power law  behavior
associated with continuous phase transitions 
or whether they show crossover behavior. For transitions
that exhibit first order characteristics, 
it would be interesting to prepare a small patch of pinning sites
with different characteristics 
that could act as a nucleation site for one of the phases 
in order to understand 
whether there is a length scale analogous to a critical nucleus size.  

We also note that the dynamics we observe should be general to 
systems with similar geometries and repulsively interacting particles. 
For example, in colloidal systems, 
square pinning arrays with flat regions
between the pinning sites (muffin-tin potentials) 
have been fabricated, and in these systems the interstitial colloids are much 
more mobile than in washboard-type pinning potentials. 
Honeycomb pinning arrays could be created using similar techniques
for this type of system.   

\section{Summary}

We have shown that vortices in honeycomb pinning arrays exhibit a rich 
variety of dynamical phases
that are distinct from those found in triangular and square pinning arrays. 
The honeycomb
pinning arrays allow for the appearance of $n$-mer type states that have 
orientational degrees of 
freedom. We specifically focused on the case where dimer states appear. 
At $B/B_\phi=2.0$, the dimers can have 
a ferromagnetic type of ordering 
which is three-fold degenerate.
At depinning, the dimers can flow in the direction 
in which they are aligned.    
For the case of driving along the $x$-axis, the dimers flow 
at $\pm 30^\circ$ to the applied drive, giving a
transverse velocity response. 
At incommensurate fields where dimers are present, 
even though the orientational order
is lost, the moving states can dynamically order into a broken 
symmetry state 
where the vortices flow with equal probability at 
either $+30^\circ$ or $-30^\circ$ to the $x$-axis.
As the driving in the $x$-direction increases, 
there is a depinning transition for the vortices in the pinning sites,
and
the transverse response is lost when the vortices either flow in a random
phase or channel along the pinning sites.
As a function of pinning force, we find 
other types of vortex lattice ordering at 
zero driving,  including a partially pinned lattice 
and a herringbone ordering of the dimers. 
These other orderings lead to new types of dynamical phases, 
including an elastic depinning for weak pinning 
where all the vortices depin simultaneously into a moving crystal phase,
and an ordered interstitial flow in which the moving dimers break apart.
The transitions between these flow phases 
appear as clear steps in the velocity force curves,
and we have mapped the dynamical phase diagrams for various system parameters. 
We also showed that the
different phases have distinct fluctuations and noise characteristics. 
When the temperature is high enough, 
the dimer states lose their orientational ordering and 
begin to rotate within the interstitial sites. 
This destroys the symmetry breaking flow; however, the
moving locked phase can still occur at high drives.    

The transition in the vortex ground state ordering as 
a function of pinning force causes
the critical depinning force for driving in the $x$ and $y$-directions
to differ. 
When driving along the $y$-direction, 
the initial depinning occurs in the form of 
one-dimensional interstitial channels, and at high drives
the vortices can form an anisotropic moving locked phase. 
We find a large enhancement of the depinning force in the $y$-direction
associated with the pinned herringbone phase
when the dimers align in the $x$-direction 
and creates a jamming effect. The jammed state can enhance the critical
depinning force by a factor of three, and 
can also arise for decreasing
pinning size.  
We expect that many of the general features we 
observe will carry over to the higher matching
fields in the honeycomb pinning arrays 
and in kagom{\' e} arrays since ordered
$n$-mers states occur for the kagom{\' e} lattice as well.

\section{Acknowledgments}
This work was carried out under the auspices of the National Nuclear
Security Administration of the U.S. Department of Energy at Los Alamos
National Laboratory under Contract No. DE-AC52-06NA25396.

\end{document}